\documentclass[12pt]{article}
\pdfoutput=1

\usepackage{hyperref}
\hypersetup{
  colorlinks=true,
  allcolors=darkpurple,
  pdfborder={0 0 0},
  linktocpage=false
}

\usepackage[normalem]{ulem}
\usepackage[utf8]{inputenc}
\usepackage[left=2.55cm, right=2.55cm, top=2.55cm, bottom=2.55cm]{geometry}
\usepackage{amsmath,amssymb}
\usepackage{slashed}
\usepackage{xcolor}
\usepackage{graphicx}
\usepackage{url}
\usepackage{cancel}
\usepackage{cite}
\usepackage{tabularx,booktabs}
\usepackage{multicol}
\usepackage{units}
\usepackage{xspace}
\usepackage[labelfont=bf]{caption}
\usepackage{mathrsfs}

\usepackage{breqn}

\definecolor{darkred}{rgb}{0.6,0,0}
\definecolor{darkpurple}{rgb}{0.5,0,0.5}

\def\gsim{\raise0.3ex\hbox{$\;>$\kern-0.75em\raise-1.1ex\hbox{$\sim\;$}}}
\def\lsim{\raise0.3ex\hbox{$\;<$\kern-0.75em\raise-1.1ex\hbox{$\sim\;$}}}

\begin{document}

\vspace*{-2cm}
\begin{flushright}
IFIC/20-39 \\
\vspace*{2mm}
\end{flushright}

\begin{center}
\vspace*{15mm}

\vspace{1cm}
{\Large \bf 
$(g-2)$ anomalies and neutrino mass } \\
\vspace{1cm}

{\bf Carolina Arbel\'aez$^{\text{a}}$, Ricardo Cepedello$^{\text{b}}$,
 Renato M. Fonseca$^{\text{c}}$, Martin Hirsch$^{\text{b}}$  }

\vspace*{.5cm} $^{\text{a}}$  Universidad T\'ecnica Federico Santa Mar\'ia
 and Centro Cient\'ifico Tecnol\'ogico\\
de Valpara\'iso CCTVal, Casilla 110-V, Valpara\'iso, Chile

\vspace*{.5cm} $^{\text{b}}$Instituto de F\'{\i}sica Corpuscular
(CSIC-Universitat de Val\`{e}ncia), \\ C/ Catedr\'atico Jos\'e
Beltr\'an 2, E-46980 Paterna (Val\`{e}ncia), Spain

\vspace*{.5cm} $^{\text{c}}$ Institute of Particle and Nuclear
Physics, Faculty of Mathematics and Physics, \\ Charles University,
V Hole\v{s}ovi\v{c}k\'ach 2, 18000 Prague 8, Czech Republic

 \vspace*{.3cm}
 \href{mailto:carolina.arbelaez@usm.cl}{carolina.arbelaez@usm.cl},
 \href{mailto:ricepe@ific.uv.es}{ricepe@ific.uv.es},
 \href{mailto:fonseca@ipnp.mff.cuni.cz}{fonseca@ipnp.mff.cuni.cz},
 \href{mailto:mahirsch@ific.uv.es}{mahirsch@ific.uv.es}
\end{center}

\vspace*{10mm}
\begin{abstract}\noindent\normalsize
  Motivated by the experimentally observed deviations from standard
  model predictions, we calculate the anomalous magnetic moments
  $a_\alpha = (g-2)_\alpha$ for $\alpha=e,\mu$ in a neutrino mass
  model originally proposed by Babu-Nandi-Tavartkiladze (BNT). We
  discuss two variants of the model, the original model plus a
  minimally extended version with an additional hypercharge zero
  triplet scalar. While the original BNT model can explain $a_\mu$,
  only the variant with the triplet scalar can explain both
  experimental anomalies. The heavy fermions of the model can be
  produced at the high-luminosity LHC and in the part of parameter
  space, where the model explains the experimental anomalies, it
  predicts certain specific decay patterns for the exotic fermions.

\end{abstract}




\section{Introduction\label{sect:intro}}

Apart from neutrino masses, as observed in oscillation experiments
\cite{Fukuda:1998mi,Ahmad:2002jz},\footnote{For a recent global fit to
  neutrino oscillation data, see for example
  \cite{deSalas:2017kay,deSalas:2020pgw}.} there are only a few
experimental hints for new physics. Among them is the long-standing
deviation of the anomalous magnetic moment of the muon from the
standard model (SM) prediction \cite{Brown:2001mga}.

Currently, the experimental data gives a roughly 4$\sigma$
c.l. deviation
\cite{Bennett:2006fi,Jegerlehner:2009ry,PDG2020,Keshavarzi:2018mgv}
from the standard model (SM) prediction:
\begin{equation}\label{eq:amu}
    \Delta a_{\mu} = (27.06 \pm 7.26) \times 10^{-10} \, .
\end{equation}  
Two new experiments will shed light on this tension: E989 experiment
at Fermilab \cite{Venanzoni:2014ixa} and E34 at J-PARC
\cite{Otani:2015jra}. E989, running since 2018, and E34, planned to
start in 2024, will improve the experimental accuracy by a factor 4
and 5, respectively, leading to a 5$\sigma$ c.l., in case the
central value of the older measurement is confirmed.

From the theory side, there is still a debate about the SM calculation
of the anomalous magnetic moment regarding hadronic vacuum
polarization (HVP). A recent lattice-QCD result
\cite{Borsanyi:2020mff} for HVP bring the SM prediction of $(g-2)$ of
the muon into agreement with experiments. However, this result is in
tension with $e^+e^- \to$ hadrons cross-section data and global fits
to electroweak precision observables \cite{Crivellin:2020zul}.

More recently, a new precise measurement of the fine-structure
constant \cite{Parker:2018vye} led to a deviation in the $(g-2)$ of
the electron \cite{Aoyama:2017uqe},
\begin{equation}\label{eq:ae}
    \Delta a_{e} = -(8.7 \pm 3.6) \times 10^{-13} \, .
\end{equation}  
Although less significant (roughly 3$\sigma$ c.l.), it provides a new
motivation to study $(g-2)$, as one might hope that both discrepancies
have a common new physics origin. While both anomalies can be
\textit{easily} explained individually, the relative sign between
$a_\mu$ and $a_e$ makes it more complicated to find a common
explanation. Simple $Z'$ (dark photon) models couple universally to
electrons and muons, and cannot account for both discrepancies
\cite{CarcamoHernandez:2019ydc}. Several papers studying both anomalies
in different contexts can be found in the literature
\cite{Crivellin:2018qmi, Dorsner:2020aaz, Liu:2018xkx, Dutta:2018fge,
  Bauer:2019gfk, Hiller:2019mou, CarcamoHernandez:2020pxw,
  Hati:2020fzp, Chen:2020jvl, Calibbi:2020emz, Botella:2020xzf,
  Chen:2020tfr, Dutta:2020scq, Jana:2020pxx}.  \\

\begin{figure}[tbph]
    \centering
    \includegraphics{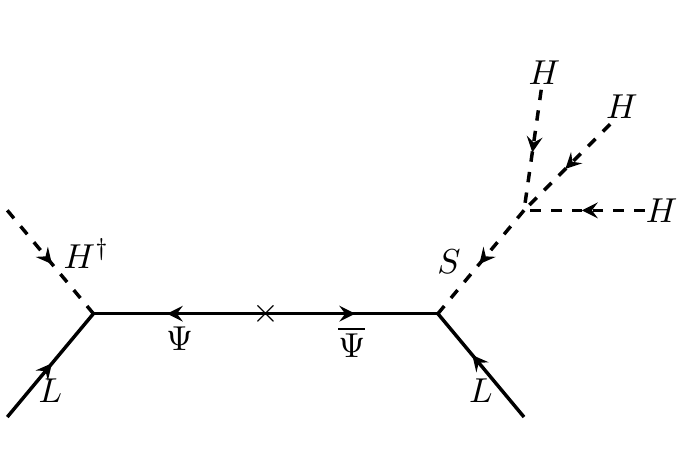}
    \caption{Dimension 7 diagram responsible for neutrino masses in
      the BNT model.}
    \label{fig:BNT-nu-masses}
\end{figure}

In this paper we study $(g-2)$ and the electric dipole moment (EDM)
for the electron and muon in the context of the
Babu-Nandi-Tavartkiladze (BNT) model \cite{Babu:2009aq} and a simple
extension of it. The BNT neutrino mass model adds to the SM particle
content vector-like fermion pairs $\left(\Psi,\overline{\Psi}\right)$,
which transform as $SU(2)_{L}$ triplets, and a scalar quadruplet
$S$. With these fields neutrino masses are induced at tree-level by a
dimension 7 operator via the diagram shown in figure
\ref{fig:BNT-nu-masses}. By closing a pair of external scalar lines, a
dimension 5 operator can also be generated with a loop.

\begin{figure}[tbph]
\begin{centering}
\includegraphics{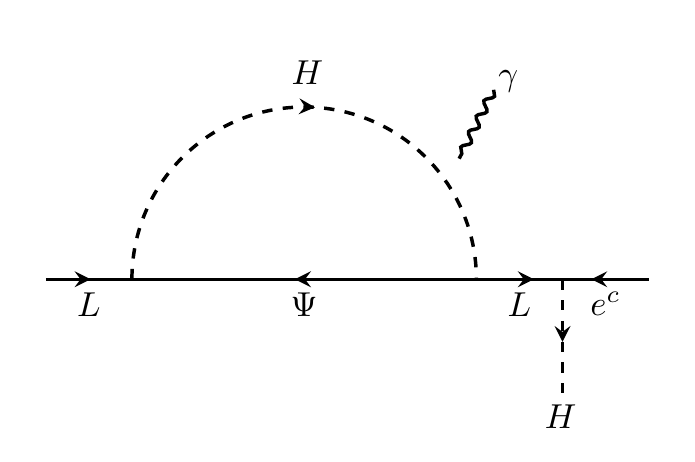}\includegraphics{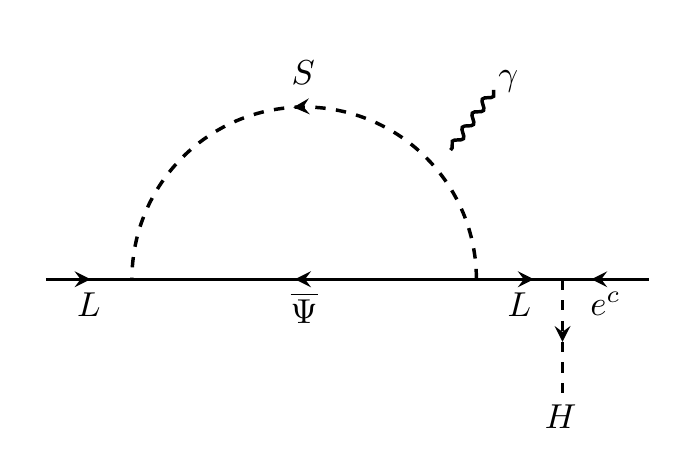}
\par\end{centering}
\caption{\label{fig:BNT-EDM}Diagrams responsible for the
  electromagnetic dipole moment operator in the BNT model. Note that
  the photon line can be attached to the internal scalar or
  fermion line.}
\end{figure}

However, with this particle content the Wilson coefficient
$c_{R}^{\alpha\beta}$ of the electromagnetic (effective) dipole moment
operator, i.e.
\begin{equation} \label{eq:op_cR}
    c_{R}^{\alpha\beta}\overline{\ell_{\alpha}}\sigma_{\mu\nu}P_{R}\ell_{\beta}F^{\mu\nu}+\textrm{h.c.} \, ,
\end{equation}
is suppressed by the small charged SM lepton masses, as can be seen
from the diagrams in figure \ref{fig:BNT-EDM}. Therefore, this model
struggles to explain the current experimentally measured value of
$\left(g-2\right)_{\mu}$, which significantly differs from the
Standard Model value. To do so very large values for Yukawa couplings
are required, which are close to the edge of non-perturbativity, given
current limits on the exotic fermion masses, see section \ref{sect:LHC}.

\begin{figure}[tbph]
\begin{centering}
\includegraphics{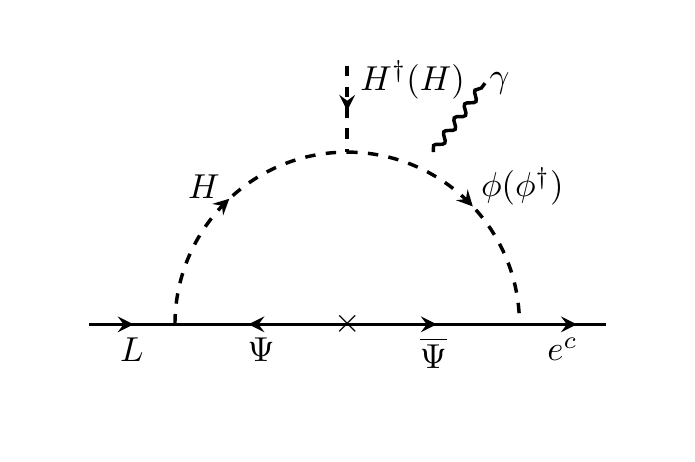}
\par\end{centering}
\caption{\label{fig:BNTphi-EDM}Main contribution to the
  electromagnetic dipole moment operator in the $\textrm{BNT}\phi$
  model. Note that the photon line can be attached to any of the internal
  scalars or fermion line.}
\end{figure}

As can be seen in eq. \eqref{eq:op_cR}, a chirality flip (mass-insertion)
is needed to close the effective operator. While the particle content
of the original BNT model requires a Higgs insertion on the SM lepton
line, the situation changes with the addition of an extra (complex)
scalar triplet $\phi$ with no hypercharge. A summary of the extra
fields in this extended BNT model --- henceforth referred to as
$\textrm{BNT}\phi$ --- can be found in table
\ref{tab:BNTphi-fields}. With this field content, the main
contribution to the electromagnetic dipole moment operator becomes
proportional to the mass of the heavy fermions
$\left(\Psi,\overline{\Psi}\right)$ rather than the muon mass, as
shown in figure \ref{fig:BNTphi-EDM} (diagram in the electroweak
basis).  

\begin{table}
\centering{}%
\begin{tabular}{ccccc}
\toprule 
 & Spin & $\mathrm{SU(3)}_{c}$ & $\mathrm{SU(2)}_{L}$ & $\mathrm{U(1)}_{Y}$\tabularnewline
\midrule
$\Psi$ & $\frac{1}{2}$ & ${\bf 1}$ & ${\bf 3}$ & $1$\tabularnewline
$\overline{\Psi}$ & $\frac{1}{2}$ & ${\bf 1}$ & ${\bf 3}$ & $-1$\tabularnewline
$S$ & 0 & ${\bf 1}$ & ${\bf 4}$ & $\frac{3}{2}$\tabularnewline
$\phi$ & 0 & ${\bf 1}$ & ${\bf 3}$ & $0$\tabularnewline
\bottomrule
\end{tabular}
\caption{\label{tab:BNTphi-fields}Quantum numbers of the new fields in
  the extended BNT model, namely $\textrm{BNT}\phi$. It contains the
  complex $\phi$ scalar which is not part of the original BNT
  model. The Weyl fermions $\Psi$ and $\overline{\Psi}$ are unrelated
  (the bar in $\overline{\Psi}$ is simply a reminder that it has the
  opposite quantum numbers of $\Psi$). One can have any number of
  generations of vector fermions $\left(\Psi,\overline{\Psi}\right)$,
  for simplicity we fix the number of copies to 3.}
\end{table}

The rest of the paper is divided as follows. In section
\ref{sect:model} we describe in detail the $\textrm{BNT}\phi$
model. We derive the neutrino masses and the relevant mixings for the
calculation of the Wilson coefficient $c_{R}^{\alpha\beta}$, directly
related to $(g-2)$, EDM and
Br($\ell_{\beta}\rightarrow\ell_{\alpha}\gamma$). At the end of the
section we derive an analytical approximation for
$c_{R}^{\alpha\beta}$, to provide some insight into the parameter
dependence of the different observables. In section \ref{sect:results}
we give and analyse the main results for the BNT and BNT$\phi$ models
to compare both. We show that while the BNT$\phi$ model can explain
both anomalous magnetic moments fulfilling the experimental
constraints, the BNT model can only account for $(g-2)_\mu$ with values
for the Yukawas at the edge of perturbativity. In section
\ref{sect:LHC} we study the phenomenology of both models at colliders
for the parameter space where they explain neutrino masses and the
anomalous magnetic moments. We finally close with a short discussion
of our results. Only the most relevant pieces of the Lagrangian are
given in the main text. The full Lagrangian can be found in the
appendix.

\section{Model setup\label{sect:model}}

\subsection{Lagrangian, masses and mixings}

We start by establishing a notation for the $SU(2)_{L}$ components
of each field. We will assume that the components of $L$ and $H$
are organized in vectors,
\begin{equation}
L=\left(\begin{array}{c}
\nu\\
\ell^{-}
\end{array}\right)\;\textrm{and}\;H=\left(\begin{array}{c}
H^{+}\\
H^{0}
\end{array}\right)\,,
\end{equation}
while the triplets $\Psi$, $\overline{\Psi}$ and $\phi$ are matrices:
\begin{equation}
\Psi=\left(\begin{array}{cc}
\frac{\Psi^{+}}{\sqrt{2}} & \Psi^{++}\\
\Psi^{0} & -\frac{\Psi^{+}}{\sqrt{2}}
\end{array}\right)\,,\;\overline{\Psi}=\left(\begin{array}{cc}
\frac{\overline{\Psi}^{-}}{\sqrt{2}} & \overline{\Psi}^{0}\\
\overline{\Psi}^{--} & -\frac{\overline{\Psi}^{-}}{\sqrt{2}}
\end{array}\right)\;\textrm{and}\;\phi=\left(\begin{array}{cc}
\frac{\phi^{0}}{\sqrt{2}} & \phi^{+}\\
\phi^{-} & -\frac{\phi^{0}}{\sqrt{2}}
\end{array}\right)\,.
\end{equation}
Lastly, $S$ is is taken to be a 3-index symmetric tensor with the
following components: $S_{111}=S^{+++}$, $S_{112}=S_{121}=S_{211}=S^{++}/\sqrt{3}$,
$S_{122}=S_{212}=S_{221}=S^{+}/\sqrt{3}$ and $S_{222}=S^{0}$. The
neutral scalars have non-zero vacuum expectation values (VEVs) which
we will denote as $\left\langle H^{0}\right\rangle \equiv v_{H}/\sqrt{2}$,
$\left\langle \phi^{0}\right\rangle \equiv v_{\phi}/\sqrt{2}$ and
$\left\langle S^{0}\right\rangle \equiv v_{S}/\sqrt{2}$. The electroweak bosons acquire masses
$m_{W}^{2}=g^{2}\left(v_{H}^{2}+4v_{\phi}^{2}+3v_{S}^{2}\right)/4$ and 
$m_{Z}^{2}=\left(g^{2}+g^{\prime2}\right)\left(v_{H}^{2}+9v_{S}^{2}\right)/4$,
therefore at tree-level the $\rho$ parameter has the value
$\left(v_{H}^{2}+4v_{\phi}^{2}+3v_{S}^{2}\right)/\left(v_{H}^{2}+9v_{S}^{2}\right)$.
In order for this number not to be far from unity,
it follows that $v_{S}$ and $v_{\phi}$ need to be much smaller
than $v_{H}$.
Indeed, assuming that only one of these two VEVs is different from zero and
using data from reference \cite{PDG2020}, the 3$\sigma$ upper limits
for $\left|v_{\phi}\right|$ and $\left|v_{S}\right|$ are roughly
4 GeV and 2 GeV, respectively.

On top of the Standard Model couplings, the $\textrm{BNT}\phi$
model contains the following mass and interaction terms:
\begin{align}
\mathcal{L}_{{\rm BNT}\phi} & =M_{\Psi}\Psi\overline{\Psi}+Y_{\Psi}L\Psi H^{*}+Y_{\overline{\Psi}}\overline{\Psi}LS+Y_{e\phi}e^{c}\overline{\Psi}\phi+Y_{e\phi^{c}}e^{c}\overline{\Psi}\phi^{*}\nonumber \\
 & +Y_{\Psi\phi}\Psi\overline{\Psi}\phi+Y_{\Psi\phi^{c}}\Psi\overline{\Psi}\phi^{*}-\mathcal{V}\,.\\
\mathcal{V} & =m_{S}^{2}S^{*}S+m_{\phi}^{2}\phi^{*}\phi+\left(\mu_{\phi}^{2}\phi\phi+\mu_{H\phi}H^{*}H\phi+\lambda_{5}S^{*}HHH\right.\nonumber \\
 & +\left.\lambda_{9}H^{*}H\phi\phi+\textrm{h.c.}\right)+\lambda_{6a}\left(H^{*}H\phi^{*}\phi\right)+\lambda_{6b}\left(H^{*}H\phi^{*}\phi\right)^{\prime}+\cdots\,.\label{eq:Potential}
\end{align}
We have omitted $SU(2)_{L}$ indices, as well as several scalar interactions
which are of little importance for this work. Nevertheless, the full
Lagrangian is displayed in appendix \ref{sec:appendix}. Flavour indices
can be read from the equations above with the understanding that the
coupling matrices have indices ordered according to the position of
the fermions; for example $Y_{\Psi}L\Psi H^{*}=\left(Y_{\Psi}\right)_{ij}L_{i}\Psi_{j}H^{*}$.

From the requirement that the first derivative of the potential is
null for the non-zero vacuum expectation values $v_{H}$, $v_{\phi}$
and $v_{S}$, together with the expected hierarchy of these VEVs,
we get the approximate tadpole equations:
\begin{align}
\mu^{2} & \approx-\frac{\lambda_{1}v_{H}^{2}}{2}+2\left(m_{\phi}^{2}+2\mu_{\phi}^{2}\right)\frac{v_{\phi}^{2}}{v_{H}^{2}} \, ,\\
\lambda_{5} & \approx-2\frac{m_{S}^{2}}{v_{H}^{3}}v_{S} \, ,\\
\mu_{H\phi} & \approx2v_{\phi}\frac{m_{\phi}^{2}+2\mu_{\phi}^{2}}{v_{H}^{2}} \, .
\label{eq:tadpole3}
\end{align}
In the very first equation, $\mu^{2}$ and $\lambda_{1}$ are the
SM scalar parameters: $\mathcal{V}_{SM}=\mu^{2}H^{*}H+\frac{1}{2}\lambda_{1}H^{*}H^{*}HH$.
Note that $\mu_{H\phi}$ is a critical parameter for the electromagnetic
dipole moment operator (see figure \ref{fig:BNTphi-EDM}) which, through
eq. \eqref{eq:tadpole3}, gets substituted by the VEV of $\phi^{0}$.
On the other hand, $\lambda_{5}$ is fundamental for the generation
of neutrino masses, as can be seen from diagram in figure \ref{fig:BNT-nu-masses},
and its value is approximately proportional to the VEV of the $S^{0}$
scalar.

\subsection{Neutrino masses}

In the basis $\left(\nu,\Psi^{0},\overline{\Psi}^{0}\right)^{T}$
the full mass matrix for neutral fermions reads, at tree-level and
in block form, 
\begin{equation}
{\cal M}^{0}=\begin{pmatrix}0 & m_{Y_{\Psi}} & m_{Y_{\overline{\Psi}}}^{T}\\
m_{Y_{\Psi}}^{T} & 0 & M_{\Psi}\\
m_{Y_{\overline{\Psi}}} & M_{\Psi}^{T} & 0
\end{pmatrix},\label{eq:Mpsi0}
\end{equation}
where $m_{Y_{\Psi}}=Y_{\Psi}v_{H}/\sqrt{2}$ and $m_{Y_{\overline{\Psi}}}=Y_{\overline{\Psi}}v_{S}/\sqrt{2}$.
With the standard seesaw approximation, if the entries in the matrices
$m_{Y_{\Psi}}\left(M_{\Psi}^{-1}\right)^{T}$ and $m_{Y_{\overline{\Psi}}}\left(M_{\Psi}^{-1}\right)^{T}$
are smaller than 1, one can block-diagonalize ${\cal M}^{0}$ and
the effective mass matrix for the light neutrinos is given by the
expression
\begin{equation}
M_{\nu}=m_{Y_{\Psi}}\left(M_{\Psi}^{-1}\right)^{T}m_{Y_{\overline{\Psi}}}+m_{Y_{\overline{\Psi}}}^{T}M_{\Psi}^{-1}m_{Y_{\Psi}}^{T}\,.\label{eq:Mnu}
\end{equation}
Note that, without loss of generality, the $M_{\Psi}$ matrix can be taken to be diagonal.

\begin{figure}[tbph]
\begin{centering}
\includegraphics{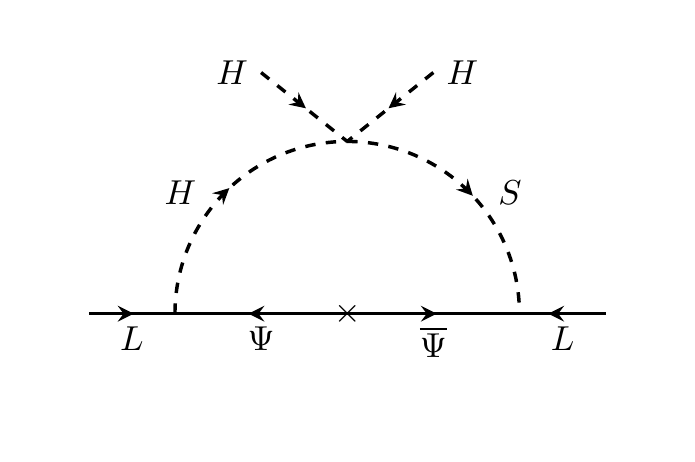}\includegraphics{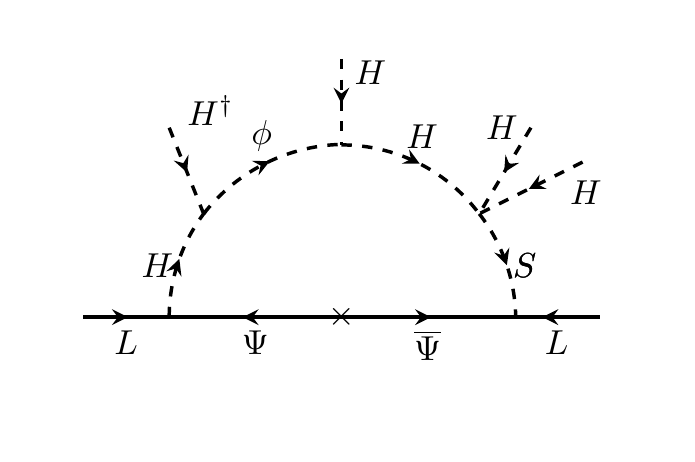}
\par\end{centering}
\caption{\label{fig:NuMass-extra-diagram}To the left: Dimension 5 diagram corresponding to the main one-loop neutrino mass contribution. To the right: Subleading dimension 7 neutrino mass contribution.}
\end{figure}

If we take the neutrino mass diagram associated to this last formula (see figure \ref{fig:BNT-nu-masses})
and close the outgoing Higgs $H^{*}$ line with one of the ingoing
Higgses $H$, we obtain also a radiative contribution to neutrino
masses already in the original BNT model. In the basis where $M_{\Psi}$
is diagonal, the correction to the tree-level formula can be expresses
as

\begin{equation}
\Delta M_{\nu}^{{\rm Loop}}=\frac{1}{16\pi^{2}}\Big(m_{Y_{\Psi}}M_{{\rm Loop}}^{-1}m_{Y_{\overline{\Psi}}}+m_{Y_{\overline{\Psi}}}^{T}M_{{\rm Loop}}^{-1}m_{Y_{\Psi}}^{T}\Big)\,,\label{eq:MnuLp}
\end{equation}
where $M_{{\rm Loop}}^{-1}$ is a diagonal matrix with entries\footnote{To a good approximation, we can use here the value of the mass of $S$ ignoring electroweak
corrections.}
\begin{equation}
\left(M_{{\rm Loop}}^{-1}\right)_{ii}\approx\frac{m_{\Psi_{i}}}{m_{S}^{2}-m_{h}^{2}}\Delta B_{0}\left(m_{S}^{2},m_{h}^{2},m_{\Psi_{i}}^{2}\right)+\frac{m_{\Psi_{i}}}{m_{S}^{2}-m_{W}^{2}}\Delta B_{0}\left(m_{S}^{2},m_{W}^{2},m_{\Psi_{i}}^{2}\right)\,.\label{eq:MnuLp-v2}
\end{equation}
The quantity $\Delta B_{0}\left(m_{A}^{2},m_{B}^{2},m_{\Psi_{i}}^{2}\right)=B_{0}\left(0,m_{A}^{2},m_{\Psi_{i}}^{2}\right)-B_{0}\left(0,m_{B}^{2},m_{\Psi_{i}}^{2}\right)$
is related to the standard Passarino-Veltman function $B_{0}$. The main contribution to the radiative neutrino mass is shown in the left panel of figure \ref{fig:NuMass-extra-diagram}. It
should be noted that, with the introduction of the $\phi$ field,
there is an extra loop contribution to neutrino masses (shown in the right panel of figure
\ref{fig:NuMass-extra-diagram}). Nevertheless, numerically its importance is small.

\subsection{Scalar masses and mixing}

In order to have a grasp on the magnitude and the parameter dependence
of the electromagnetic dipole moment operator in expression \eqref{eq:op_cR},
it is important to understand how do the various scalars mix. That
discussion is simplified if we consider that the VEV of $S$ is negligible
($v_{S}\approx0$), in which case this field does not mix with the
remaining ones. All scalar components of $S$ can therefore be safely
ignored (except in what concerns neutrino mass generation).
However, we note that while this and other approximations made in the following discussion are very usefulness,
to obtain the numerical results in section \ref{sect:results} we used the full un-approximated expressions
for masses and mixing angles.

We are left with three scalars with electric charge +1 contributing
for the dipole moment operator, namely $H^{+}$, $\phi^{+}$ and $\left(\phi^{-}\right)^{*}$.
For this particular ordering of the fields, their mass-squared matrix
is given by the approximate expression
\begin{align}
{\cal M}^{+} & \approx\left(\begin{array}{ccc}
2\frac{v_{\phi}^{2}}{v_{H}^{2}}\left(\kappa_{1}+\kappa_{2}\right) & \frac{\sqrt{2}v_{\phi}}{v_{H}}\kappa_{1} & \frac{\sqrt{2}v_{\phi}}{v_{H}}\kappa_{2}\\
\frac{\sqrt{2}v_{\phi}}{v_{H}}\kappa_{1} & \kappa_{3} & \kappa_{1}-\kappa_{3}\\
\frac{\sqrt{2}v_{\phi}}{v_{H}}\kappa_{2} & \kappa_{1}-\kappa_{3} & -\kappa_{1}+\kappa_{2}+\kappa_{3}
\end{array}\right)\,,
\end{align}
with
\begin{align}
\kappa_{1} & \approx m_{\phi}^{2}+2\mu_{\phi}^{2}+\left(\lambda_{6a}+\lambda_{6b}+2\lambda_{9}\right)\frac{v_{H}^{2}}{2}\,,\\
\kappa_{2} & \approx \kappa_{1}-\lambda_{6b}\frac{v_{H}^{2}}{2}\,,\\
\kappa_{3} & \approx m_{\phi}^{2}+\left(\lambda_{6a}+\lambda_{6b}\right)\frac{v_{H}^{2}}{2}\,.
\end{align}
The properly normalized admixture of fields $v_{H}H^{+}-\sqrt{2}v_{\phi}\phi^{+}-\sqrt{2}v_{\phi}\left(\phi^{-}\right)^{*}$
constitutes the pseudo-Goldstone boson $G^{+}$, therefore we are
left with only two other mass eigenstates: $\varphi_{1}^{+}$ and
$\varphi_{2}^{+}$. In order to make a quantitative analysis of their
masses and composition, we shall take into account that $v_{\phi}$
must be significantly smaller than $v_{H}$, which in turn is much
smaller than the bare masses $m_{\phi}^{2}$ and $\mu_{\phi}^{2}$.
When this is the case,
\begin{align}
m^{2}\left(\varphi_{1}^{+}\right) & \approx m_{\phi}^{2}+2\mu_{\phi}^{2}+\left(\lambda_{6a}+\frac{1}{2}\lambda_{6b}+2\lambda_{9}\right)\frac{v_{H}^{2}}{2}\,,\\
m^{2}\left(\varphi_{2}^{+}\right) & \approx m_{\phi}^{2}-2\mu_{\phi}^{2}+\left(\lambda_{6a}+\frac{1}{2}\lambda_{6b}-2\lambda_{9}\right)\frac{v_{H}^{2}}{2}\,.
\end{align}
There is also the following relation between electroweak and mass
eigenstates:
\begin{align}
\left(\begin{array}{c}
G^{+}\\
\varphi_{1}^{+}\\
\varphi_{2}^{+}
\end{array}\right) & \approx\left(\begin{array}{ccc}
1 & -\sqrt{2}\frac{v_{\phi}}{v_{H}} & -\sqrt{2}\frac{v_{\phi}}{v_{H}}\\
2\frac{v_{\phi}}{v_{H}} & \frac{1}{\sqrt{2}} & \frac{1}{\sqrt{2}}\\
0 & -\frac{1}{\sqrt{2}} & \frac{1}{\sqrt{2}}
\end{array}\right)\left(\begin{array}{c}
H^{+}\\
\phi^{+}\\
\left(\phi^{-}\right)^{*}
\end{array}\right)\,.
\end{align}

The neutral scalars, $H^{0}$ and $\phi^{0}$, are equally important
in the loop diagrams \ref{fig:BNT-EDM}. Splitting these two fields
in their real and imaginary parts, 
\begin{equation}
H^{0}\equiv\frac{H_{R}^{0}+iH_{I}^{0}}{\sqrt{2}} \, , \quad \phi^{0}\equiv\frac{\phi_{R}^{0}+i\phi_{I}^{0}}{\sqrt{2}},
\end{equation}
it is easily seen that only the CP-even fields $H_{R}^{0}$ and $\phi_{R}^{0}$
mix to form two mass eigenstates. One of them should be identified
with the observed 125 GeV Higgs boson and the other one we will call
$R^{0}$. The CP-odd fields --- $H_{I}^{0}$ and $\phi_{I}^{0}$
--- do not mix, hence they are mass eigenstates. In particular, $H_{I}^{0}\equiv G^{0}$
is the neutral pseudo-Goldstone boson, just like in the Standard Model.
The expressions for the mass eigenstates are
\begin{equation}
h^{0}\approx H_{R}^{0}+2\frac{v_{\phi}}{v_{H}}\phi_{R}^{0},
\quad 
R^{0}\approx\phi_{R}^{0}-2\frac{v_{\phi}}{v_{H}}H_{R}^{0},
\quad
G^{0}=H_{I}^{0},
\quad
\phi_{I}^{0}\,,
\end{equation}
with the corresponding (pseudo)masses given by the following formulas:
\begin{equation} 
    m^{2}\left(h^{0}\right) \approx 4 \lambda_{1}v_{H}^{2}-4\left(m_{\phi}^{2}+2\mu_{\phi}^{2}\right)\frac{v_{\phi}^{2}}{v_{H}^{2}} \, ,
\quad
m^{2}\left(R^{0}\right) \approx m^{2}\left(\varphi_{1}^{+}\right) \, ,
\end{equation}
\begin{equation} 
    m^{2}\left(G^{0}\right) =0 \, , \quad m^{2}\left(\phi_{I}^{0}\right)  \approx m^{2}\left(\varphi_{1}^{-}\right) \, .
\end{equation}

\subsection{Analytical understanding of the value of $\left(g-2\right)_{e,\mu}$
and related observables}

With the approximate dependence of masses and mixing angles on the
various Lagrangian parameters, we are in a position to estimate the
value of $g-2$ for the electron and the muon, as well as the value
of the lepton electric dipole moments and the branching ratios $\textrm{Br}\left(\ell_{\beta}\rightarrow\ell_{\alpha}\gamma\right)$.
Nevertheless, for our numerical results we used the full one-loop expressions,
without approximations.

Following \cite{Crivellin:2018qmi}, we can describe a generic interaction
of a fermion $\Psi$ with the charged leptons $\ell_{\alpha}=e,\mu,\tau$
and a scalar $\Phi$ with couplings $\Gamma_{L}^{\alpha}$ and $\Gamma_{R}^{\alpha}$:
\begin{equation}
\mathscr{L}=\cdots+\overline{\Psi}\left(\Gamma_{L}^{\alpha}P_{L}+\Gamma_{R}^{\alpha}P_{R}\right)\ell_{\alpha}\Phi^{*}+\textrm{h.c.}
\end{equation}
Through a loop diagram, this interaction will induce a contribution
to the electromagnetic dipole moment operator shown in eq. \eqref{eq:op_cR}.
Specifically, reference \cite{Crivellin:2018qmi} quotes the result
\begin{equation}
c_{R}^{\alpha\beta}\approx\frac{1}{16\pi^{2}}\Gamma_{L}^{\alpha*}\Gamma_{R}^{\beta}\frac{m_{\Psi}}{m_{\Phi}^{2}}\left[f_{\Phi}\left(\frac{m_{\Psi}^{2}}{m_{\Phi}^{2}}\right)+q_{\Psi}g_{\Phi}\left(\frac{m_{\Psi}^{2}}{m_{\Phi}^{2}}\right)\right]
\end{equation}
plus sub-leading contributions which are suppressed by the masses
of the leptons $\ell_{\alpha}$ and $\ell_{\beta}$. The loop functions
appearing in this expression are
\begin{align}
f\left(x\right) & =\frac{x^{2}-1-2x\log x}{4\left(x-1\right)^{3}}\,,\\
g\left(x\right) & =\frac{x-1-\log x}{2\left(x-1\right)^{2}}\,,
\end{align}
and $q_{\Psi}$ is the electric charge of the loop fermion, flowing
from the initial lepton ($\ell_{\beta}$) to the final one ($\ell_{\alpha}$).

The lepton anomalous magnetic moments, their electric dipole moments,
as well as the branching ratios of the decays $\ell_{\beta}\rightarrow\ell_{\alpha}\gamma$
can readily be calculated from the numbers $c_{R}^{\alpha\beta}$:
\begin{align}
\frac{\left(g-2\right)_{\alpha}}{2}\equiv a_{\alpha} & =-4\frac{m_{\ell_{\alpha}}}{e} \, \textrm{Re} \, c_{R}^{\alpha\alpha}\,,\\
d_{\alpha} & =-2 \, \textrm{Im} \, c_{R}^{\alpha\alpha}\,,\\
\textrm{Br}\left(\ell_{\beta}\rightarrow\ell_{\alpha}\gamma\right) & =\frac{m_{\ell_{\beta}}^{3}}{4\pi\Gamma_{\ell_{\beta}}}\left(\left|c_{R}^{\alpha\beta}\right|^{2}+\left|c_{R}^{\beta\alpha}\right|^{2}\right)\,.
\end{align}
For the particular model under discussion, the main contribution to
$c_{R}^{\alpha\beta}$ is due to the loop shown in figure \ref{fig:BNTphi-EDM}.
With the approximate value of the scalar mixing matrices provided
earlier, we obtain the following estimate:
\begin{align} \label{eq:coeffapprox}
c_{R}^{\alpha\beta} & \approx\frac{1}{16\pi^{2}}\sum_{i=1}^{3}\left(Y_{\Psi}\right)_{\alpha i}\left(Y_{e\phi}+Y_{e\phi^c}\right)_{\beta i}m_{\Psi_{i}}\frac{v_{\phi}}{v_{H}}\nonumber \\
 & \times\sum_{\Phi=\left\{ G^{+},h^{0},\varphi_{1}^{+},R^{0}\right\} }\frac{\kappa_{\Phi}}{m_{\Phi}^{2}}\left[f\left(\frac{m_{\Psi_{i}}^{2}}{m_{\Phi}^{2}}\right)-\kappa_{\Phi}^{\prime}g\left(\frac{m_{\Psi_{i}}^{2}}{m_{\Phi}^{2}}\right)\right]\,.
\end{align}
The $\kappa_{\Phi}^{(\prime)}$ coefficients in this expression take
the following values: $\kappa_{G^{+},h^{0},\varphi_{1}^{+},R^{0}}=\left(-\sqrt{2},1,\sqrt{2},-1\right)$
and $\kappa_{G^{+},h^{0},\varphi_{1}^{+},R^{0}}^{\prime}=\left(2,1,2,1\right)$.
At leading order, the remaining scalars do not contribute. Note also
that the present model does not contain extra gauge bosons, however
the couplings of the Standard Model ones to leptons is slightly altered.
Instead of considering an extra loop with an internal $W^{\pm}$ boson,
it is sufficient to include in the scalar computation the pseudo-Goldstone
boson $G^{+}$, assigning to it its physical mass ($m_{G^{+}}=m_{W}$).
\section{Results and discussion\label{sect:results}}

In this section we will discuss the numerical results for $\Delta
(g-2)_{\alpha}$ ($\alpha=e,\mu$), the neutrino mass fits and
constraints from charged lepton flavour violation (cLFV) searches. We
also briefly comment on electric dipole moments, $d_{e,\mu}$. The
discussion is divided into two parts. In the first one, we show
results for the original BNT model, while the second discusses the
extended version, BNT$\phi$.

We have implemented the model into \texttt{SARAH}
\cite{Staub:2012pb,Staub:2013tta}. This program generates \texttt{SPheno}
routines \cite{Porod:2003um,Porod:2011nf} for a numerical calculation
of mass spectra and other observables. To cross-check the results,
we have written private codes for numerical evaluation
of $\Delta (g-2)_{\alpha}$ and implemented also the approximation
formulas discussed in \ref{sect:model}.

\subsection{Results for the BNT model\label{subsect:RBNT}}

Since the main motivation for the BNT model \cite{Babu:2009aq} is
to explain the observed neutrino oscillation data, we first briefly
discuss, how neutrino masses and angles can be fitted to the
experimental data. 

The master parametrization
\cite{Cordero-Carrion:2018xre,Cordero-Carrion:2019qtu} allows us to
fit any Majorana neutrino mass model to experimental data.
For the case of the BNT model, the general formulae in
\cite{Cordero-Carrion:2019qtu} simplify to

\begin{eqnarray}\label{eq:master}
  Y_1 = c M_{\Psi}^{1/2} W T {\hat m_{\nu}}^{1/2} U_{\nu}^{\dagger} \, , \\ \nonumber
  Y_2 = c M_{\Psi}^{1/2} W^* B {\hat m_{\nu}}^{1/2} U_{\nu}^{\dagger} \, ,
\end{eqnarray}
with
\begin{equation}\label{eq:Bmat}
  B = (T^T)^{-1} (\mathbb{I} - K)  \, ,
\end{equation}
and $c=(v_H v_S)^{-1/2}$. Since the neutrino mass matrix is symmetric
under the exchange of $Y_1$ and $Y_2$, we can associate $Y_{\Psi}$ and
$Y_{\overline\Psi}$ with either of them arbitrarily.  The master
parametrization calculates the two Yukawa matrices as function of the
input parameters, $m_{\nu_i}$, $U_{\nu}$ and $M_{\Psi}$ and three
matrices, $W$, $T$ and $K$ with arbitrary parameters. Here, $W$ is a
unitary, $T$ an upper triangular and $K$ an antisymmetric matrix. All
matrices are $(3,3)$. As usual, ${\hat m_{\nu}}$ and $U_{\nu}$ are the
light neutrino mass eigenvalues and mixing matrix.
\footnote{An alternative, but equivalent fit could be done
  using one of the two Yukawa couplings as input:
\begin{equation}\label{eq:FitR}
Y_{\Psi} = \Big(\frac{1}{v_H v_S} M_{\nu} + A \Big) (Y_{\overline\Psi})^{-1}M_\Psi,
\end{equation}  
with $A$ being a generic anti-symmetric matrix and $M_{\nu}$ the
neutrino mass matrix in the flavour basis.}

\begin{figure}[t]
\centering
\includegraphics[width=0.45\textwidth]{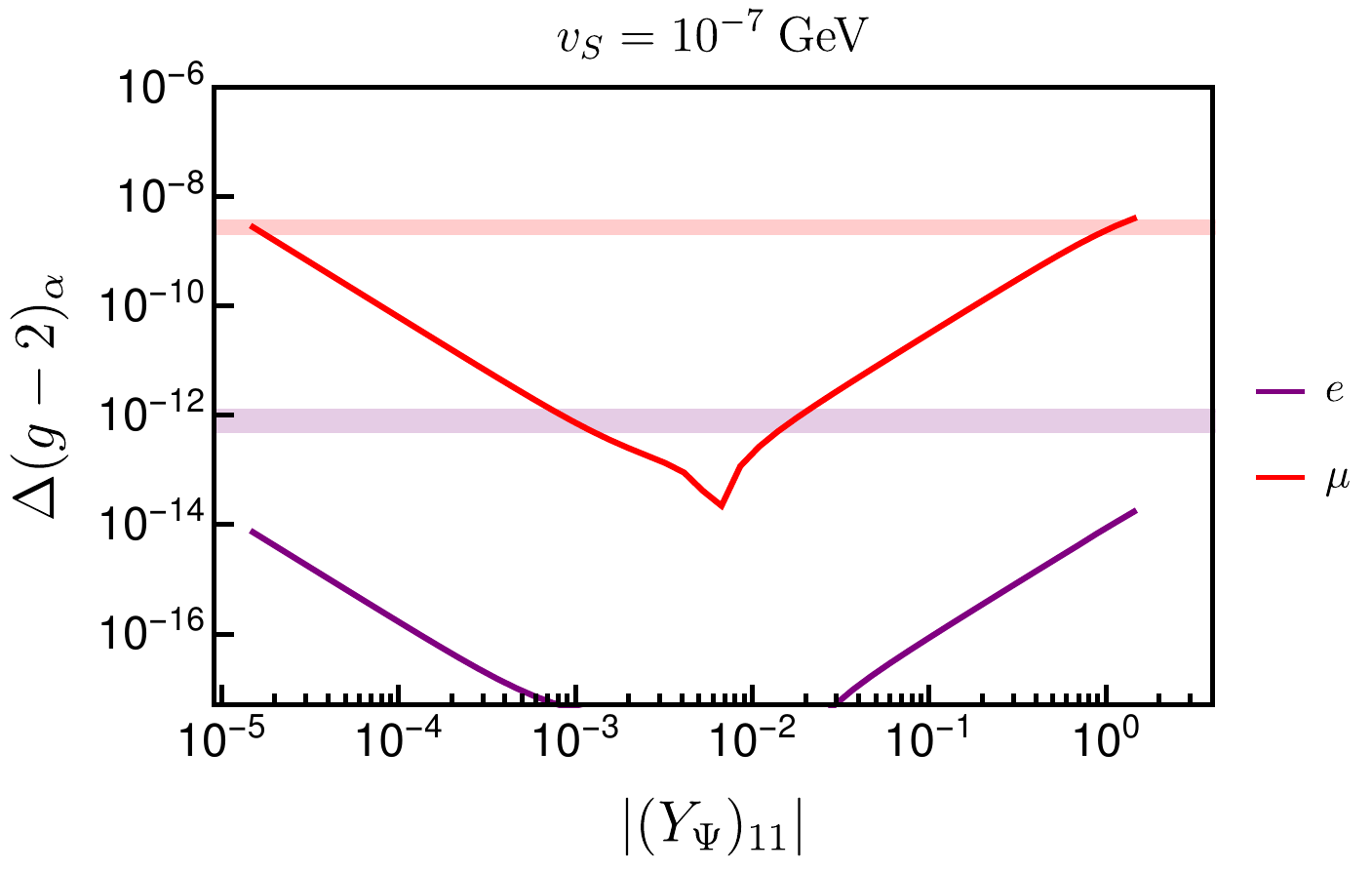}
\includegraphics[width=0.52\textwidth]{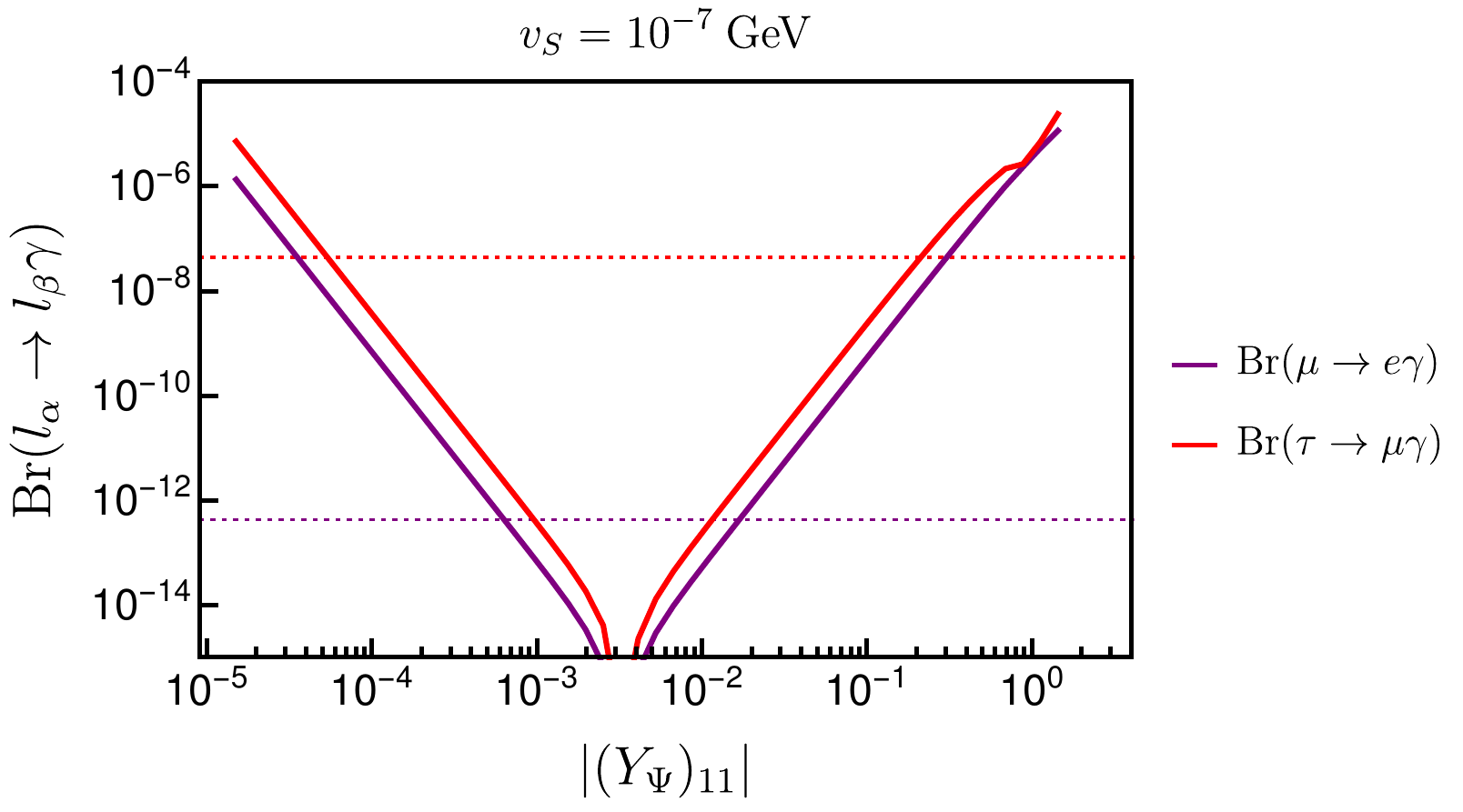}
\includegraphics[width=0.45\textwidth]{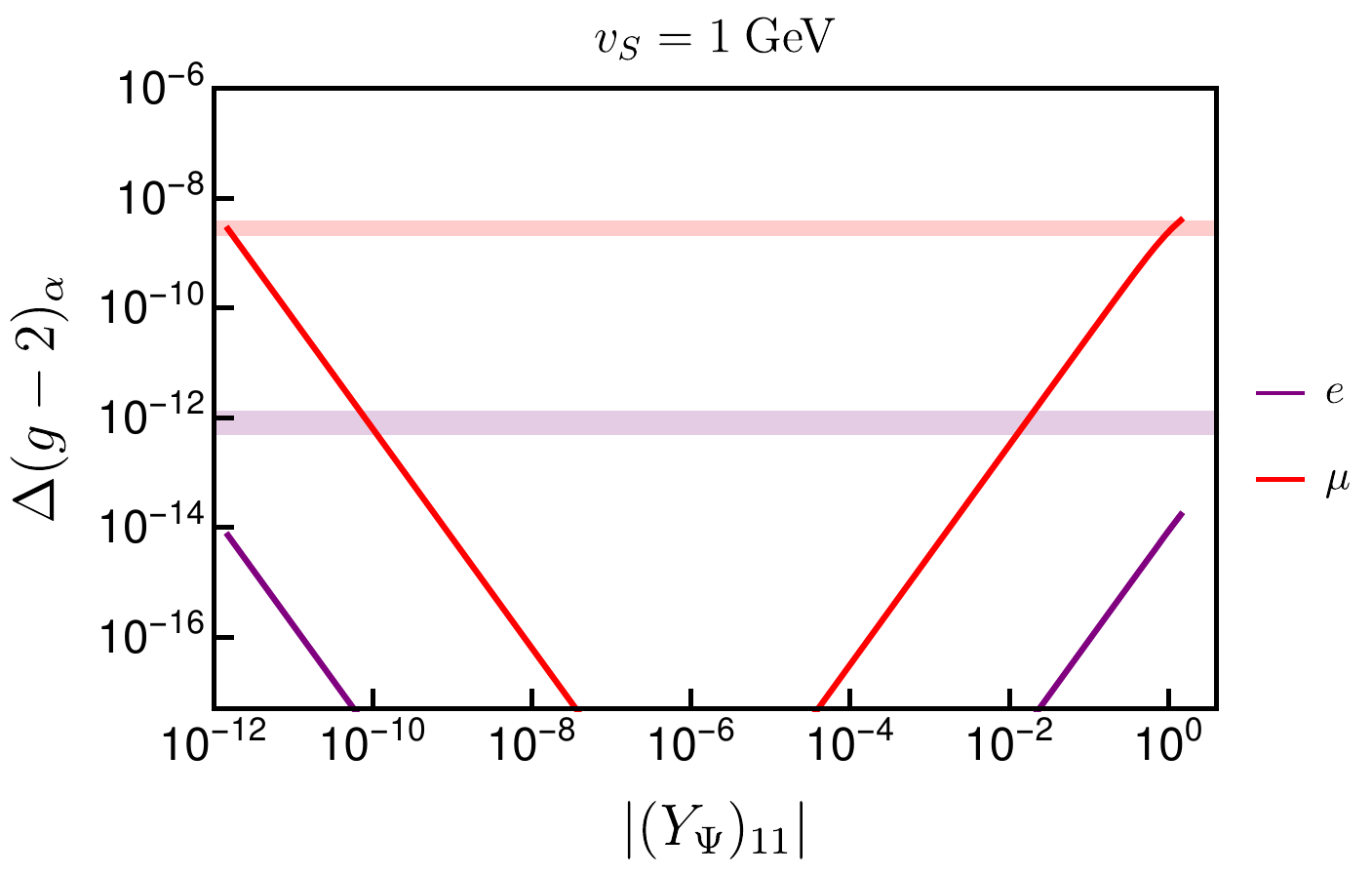}
\includegraphics[width=0.52\textwidth]{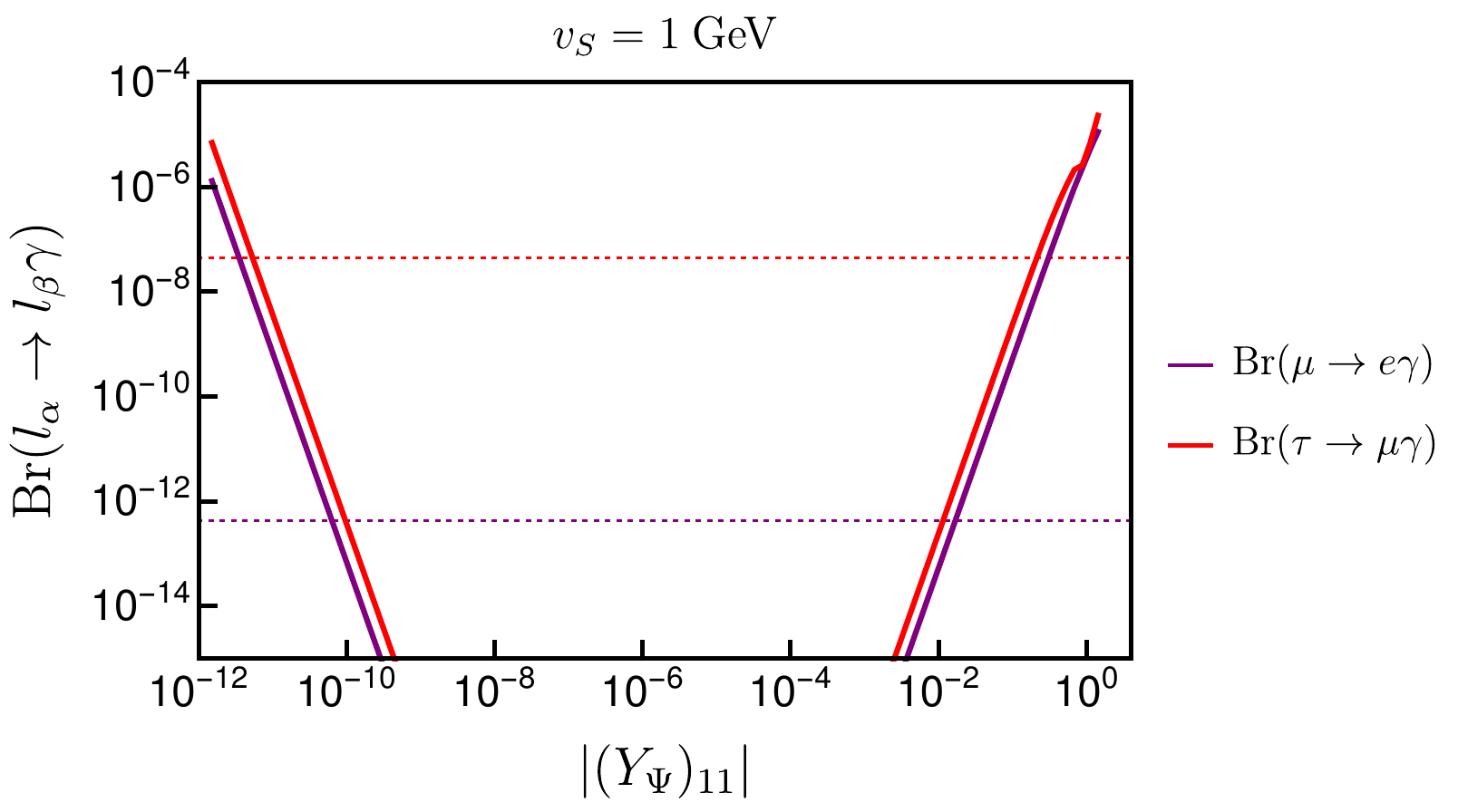}
\caption{Left: $\Delta(g-2)_{\alpha}$, the shift with respect to the
  standard model value of $(g-2)_{\alpha}$, as a function of
  $|(Y_{\Psi})_{11}|$ for 2 choices of the quadruplet VEV, $v_S$.
  Shown are only $\alpha=e,\mu$, since for the $\tau$ there is
  practically no experimental information. The thin horizontal bands
  are the 1 $\sigma$ c.l. ranges of the experimental anomalies.
  Right: Br$(l_i\to l_j\gamma$) for the same choice of parameters as
  the left figures. The masses of $\Psi_i$ are chosen 
  with $m_{\Psi_{1,2,3}}=(0.8,0.9,1.0)$ TeV, the mass of the scalar quadruplet is
  $m_{S}=500$ GeV. For the neutrino fit, see the discussion in the text. 
    \label{BNTgm2}}
\end{figure}

An especially simple case is the choice $W=\mathbb{I}$,
$T=f\times\mathbb{I}$ and $K=0$.  This leads to $Y_{\Psi}=f^2
Y_{\overline\Psi}^T$ and both Yukawa matrices are equal for $f=1$.  With
this choice, both Yukawa matrices have off-diagonal elements, due to
the large mixing angles, observed in oscillation experiments.
Fig.(\ref{BNTgm2}) shows $\Delta(g-2)_{\alpha}$ and Br$(l_i\to
l_j\gamma)$ for some fixed masses $m_{\Psi_i}$ as function of the
diagonal entries in $Y_{\Psi}$, $|(Y_{\Psi})_{ii}|$. Since the neutrino
fit requires the product of the two Yukawa matrices to be constant,
small values of $|(Y_{\Psi})_{ii}|$ correspond to large entries in
$Y_{\overline\Psi}$ and vice versa.

As the plots in fig.(\ref{BNTgm2}) show $\Delta(g-2)_{\mu}$ can be
explained if either $|(Y_{\Psi})_{22}|$ or
$|(Y_{\overline\Psi})_{22}|$ are order ${\cal O}(1)$. However,
$\Delta(g-2)_{e}$ is always smaller than the experimental
anomaly.\footnote{Since we plot logarithmically, the plot shows
  $|\Delta(g-2)_{e}|$. Experimentally $\Delta(g-2)_{\mu}$ and
  $\Delta(g-2)_{e}$ have different signs. We have checked that the
  relative signs can be easily generated, by relative signs in the
  entries of the Yukawa matrices.} Even more importantly, cLFV
constraints, especially Br$(\mu\to e \gamma)$ rule out all points,
which explain $\Delta(g-2)_{\mu}$ in this fit.

\begin{figure}[t]
\centering
\includegraphics[width=0.48\textwidth]{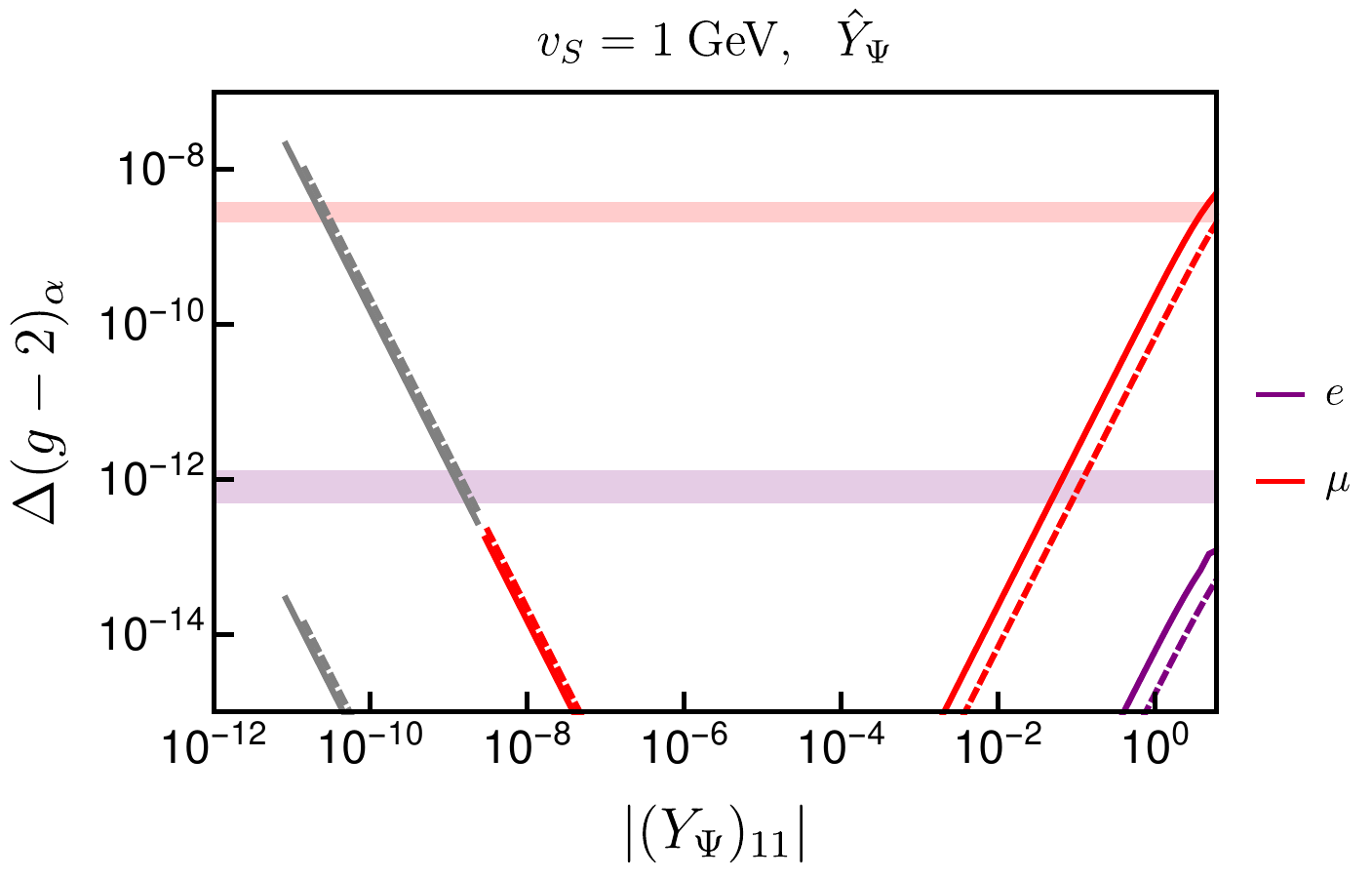}
\includegraphics[width=0.48\textwidth]{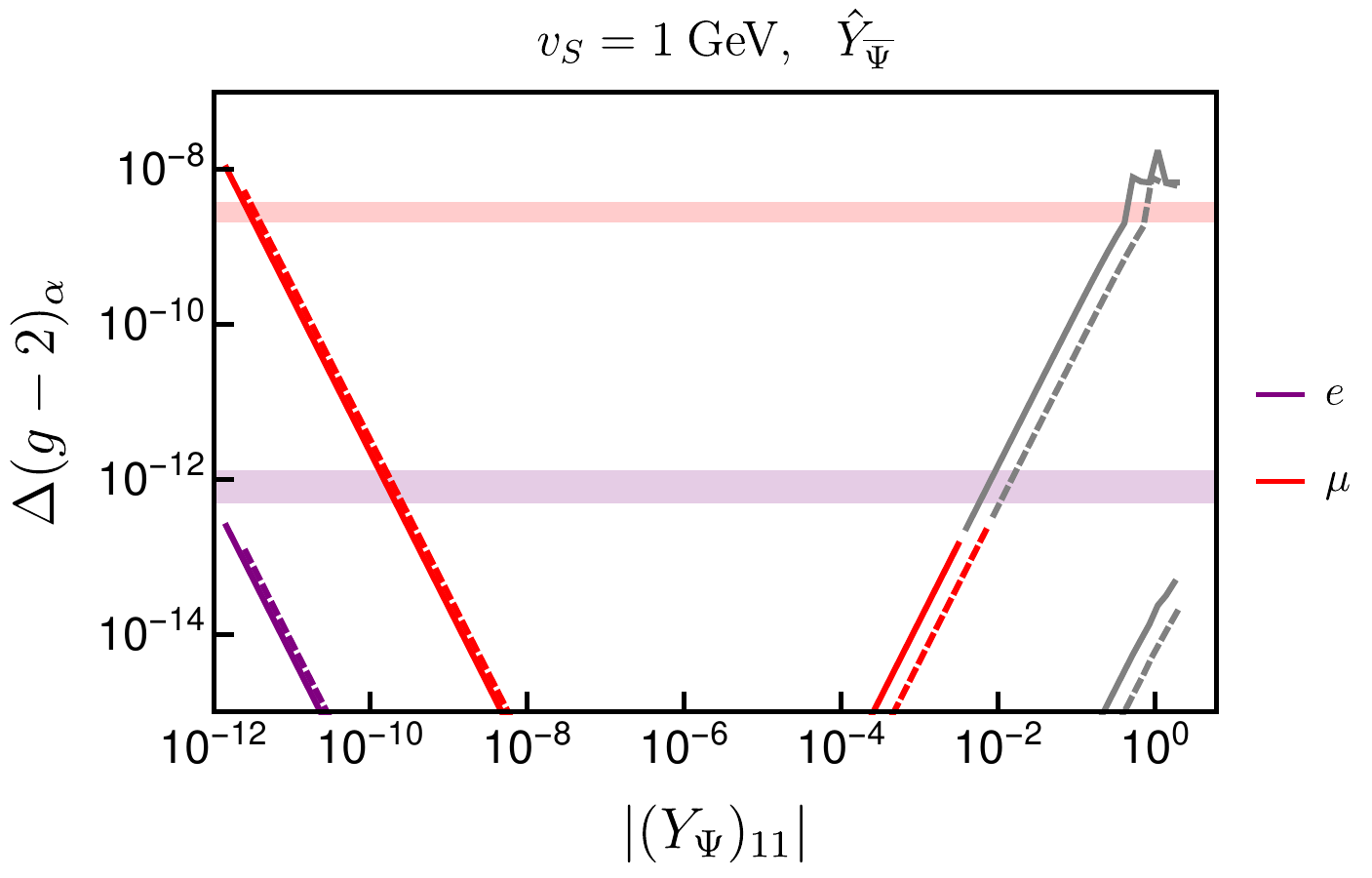}
\caption{$\Delta(g-2)_{\alpha}$ as a function of $|(Y_{\Psi})_{11}|$.
  $m_{\Psi_i}$ are chosen $m_{\Psi_{1,2,3}}=(0.8,0.9,1.0)$ TeV (full
  lines) and $m_{\Psi_{1,2,3}}=(1.5,1.6,1.7)$ TeV (dashed).  Grey
  points are ruled out by the experimental limit on Br$(\mu\to e
  \gamma)$.  For the neutrino fit, see the discussion in the text.
    \label{BNTgm2a}}
\end{figure}

We have therefore tried a different ansatz for the matrices $W$ and
$T$. It is easy to show that the choice $W=U_{\nu}$ and $T=f\times
{\hat m_{\nu}}^{-1/2}$ will lead to a fit of neutrino data in which
one of the two Yukawa matrices is diagonal.  Fig.(\ref{BNTgm2a}) shows
the result of this calculation. Here we show $\Delta(g-2)_{\mu}$ for
two choices of $m_{\Psi}$. Full lines are for $m_{\Psi}=800$ GeV,
dashed lines for $m_{\Psi}=1.5$ TeV. These values are motivated by the
(estimated) lower limit and future sensitivity of the LHC, see the
discussion in section \ref{sect:LHC}. Points in colour are allowed by
cLFV constraints, while points violating the experimental bound on
Br$(\mu\to e \gamma)$ are shown in grey.  The plot to the left fits
neutrino data with a diagonal matrix $Y_{\Psi}$, while the plot to the
right is for diagonal $Y_{\overline\Psi}$. As expected, the model can
explain $\Delta(g-2)_{\mu}$, consistent with the bound on
Br$(\mu\to e \gamma)$, if the larger of the two Yukawa matrices is
diagonal. The plots show that both, ${\hat Y_{\Psi}}$ and
${\hat Y_{\overline\Psi}}$ give valid solutions. Again, $\Delta(g-2)_{e}$
is never large enough to explain the experimental anomaly.

\begin{figure}[t]
\centering
\includegraphics[width=0.65\textwidth]{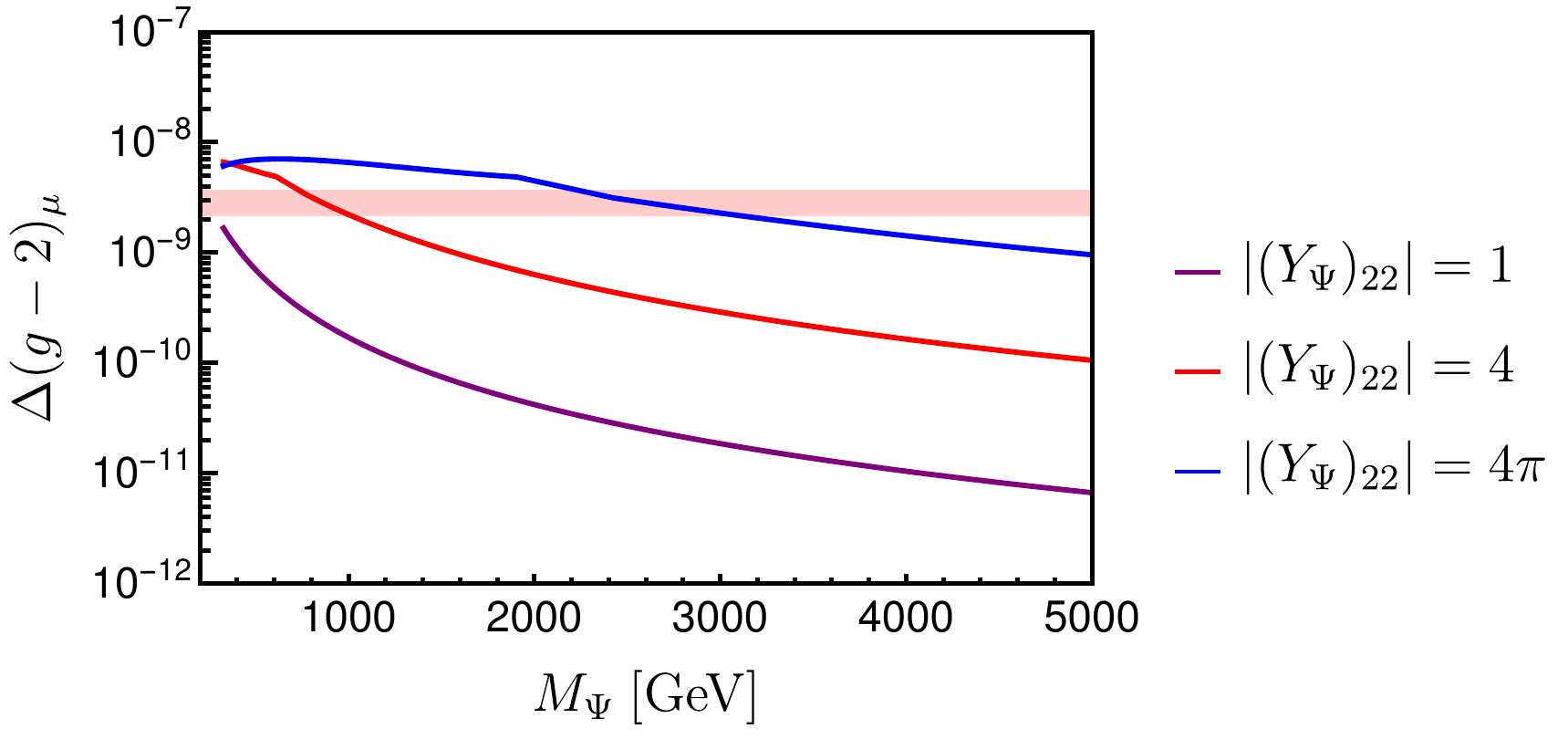}
\caption{$\Delta(g-2)_{\mu}$ for 3 fixed values of $(Y_{\Psi})_{22}$
  as a function of $M_{\Psi}$, for degenerate $\Psi$. 
    \label{BNTgm2mu}
  }
\end{figure}

While the BNT model can explain $\Delta(g-2)_{\mu}$, at least one
of the Yukawas needs to be ${\cal O}(1)$, given current lower limits
on the heavy fermion masses. In fact, as we show in figure \ref{BNTgm2mu},
in the BNT model one can derive an upper bound on the mass of
$\Psi$, from the requirement that the experimental anomaly is correctly
explained. As figure \ref{BNTgm2mu} demonstrates, even for
$|(Y_{\Psi})_{22}|=4\pi$, $m_{\Psi}$ can not be larger than roughly
3 TeV in this case. Note that such a large coupling makes the
model non-perturbative, thus this number is conservative and the
LHC should be able to test values up to $|(Y_{\Psi})_{22}| \sim (6-7)$.

\subsection{Results for the BNT$\phi$ model\label{subsect:RBNTphi}}

\begin{figure}[t]
\centering
\includegraphics[width=0.49\textwidth]{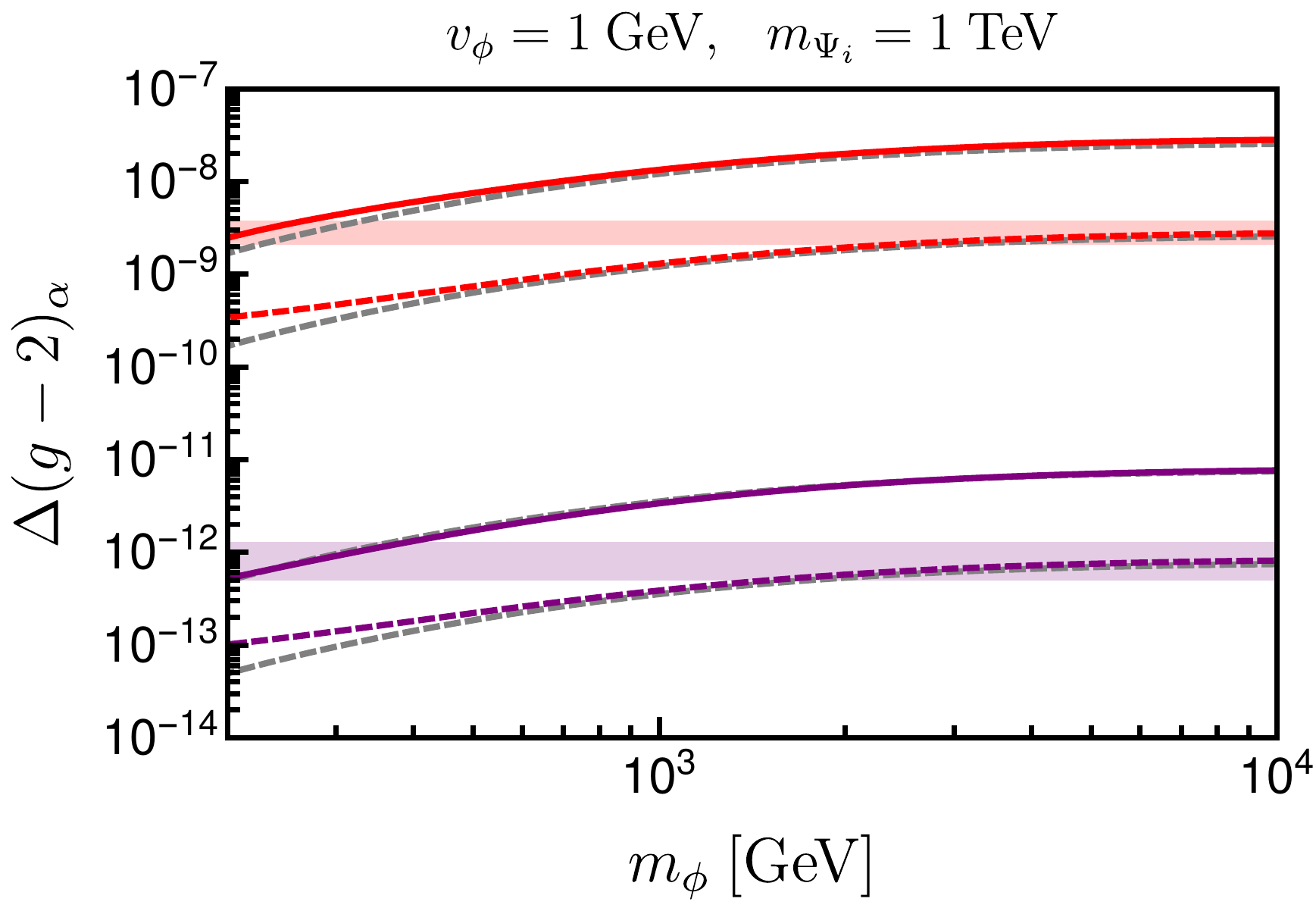}
\includegraphics[width=0.49\textwidth]{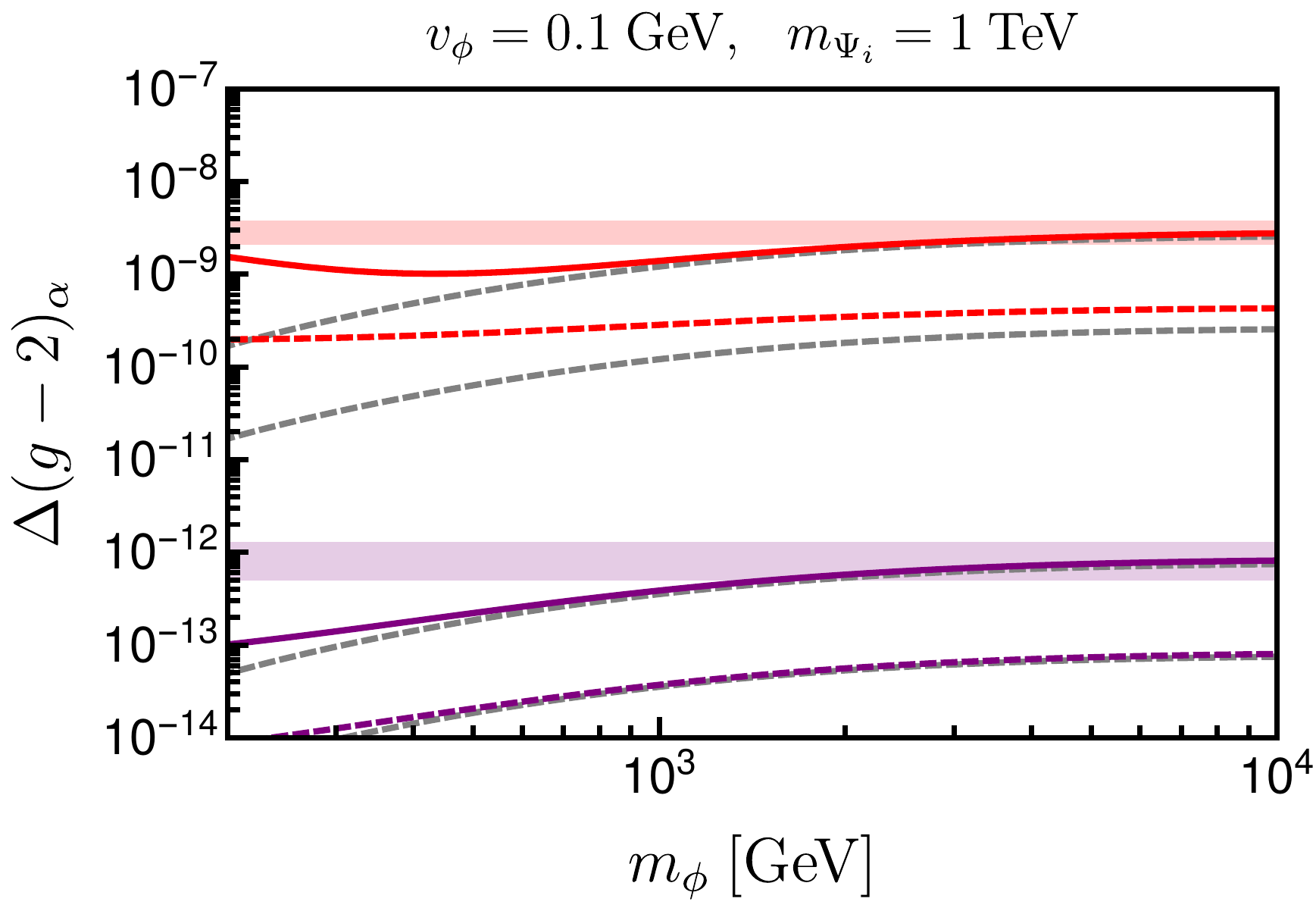}
\caption{$\Delta(g-2)_{\alpha}$ as a function of $m_{\phi}$  
  for different values of $v_{\phi}$ and $Y_{e\phi}=Y_{e\phi^c}$.
  The coloured lines are the numerical result, the gray lines
  are calculated with the approximation formulas, see text.
  Full lines are calculated with $(Y_{e\phi})_{11}=0.06$ and
  $(Y_{e\phi})_{22}=1$, dashed lines  $(Y_{e\phi})_{11}=0.006$ and
  $(Y_{e\phi})_{22}=0.1$. As in all other figures in this section,
  red is for $\mu$ and purple for $e$.
    \label{BNTphigm2}}
\end{figure}

Let us turn now to a discussion of the extended BNT model.
The addition of the scalar $\phi$ to the model generates diagrams,
for which the mass flip, necessary for the generation of
$\Delta(g-2)_{\alpha}$, can be caused by the large fermion mass 
internal to the loop.

Figure \ref{BNTphigm2} shows a comparison of the approximation
formulas, see section \ref{sect:model}, to the full numerical results
from \texttt{SPheno}.  The plot shows $\Delta(g-2)_{\alpha}$ as a
function of $m_{\phi}$ for different values of $v_{\phi}$ and
$Y_{e\phi}=Y_{e\phi^c}$.  The neutrino fit was done with a diagonal
coupling $Y_{\Psi}$, with entries on the diagonal equal to
$(Y_{\Psi})_{ii}=1$ for simplicity. The plots demonstrate that there
is a large range of parameter space, for which both experimental
anomalies can be explained simultaneously. It may seem counter-intuitive
that $\Delta(g-2)_{\alpha}$ rises with increasing mass $m_{\phi}$. The
reason for this is the relative sign between the diagrams from
the Goldstones and the scalars, see eq. \eqref{eq:coeffapprox}. 
This sign leads to a cancellation in  $\Delta(g-2)_{\alpha}$ if the
scalars $h^0$ and $\phi^+$ are degenerate with the corresponding
Goldstone bosons. For large $m_{\phi}$ this cancellation is less effective
and in the limit $m_{\phi}\to\infty$, only the Goldstone diagrams
contribute to the observable.

Figure \ref{BNTphigm2} also demonstrates that the approximation
formulas work quite well for $v_{\phi}$ order ${\cal O}({\rm GeV})$,
as expected. For the muon, the approximation formula starts to
differ from the numerical results, once $v_{\phi}$ and
$Y_{e\phi}$ are smaller than GeV and $Y_{e\phi}\ll 1$, respectively,
while for the electron the approximation still works reasonably.
Again, this is to be expected, since $m_{\mu}/m_e \simeq 200$,
such that diagrams with external mass flips are more important
in the case of the muon. Depending on other model parameters,
the diagrams proportional to  $v_{\phi}$ will be sub-dominant
even for very large (i.e. non-perturbative) Yukawa couplings
for  $v_{\phi} \lsim (10^{-3}-10^{-2})$ GeV. For $v_{\phi}$ below
$10^{-3}$ GeV the model can not explain $\Delta(g-2)_{e}$ and
the results of the original BNT model are approximately recovered.

\begin{figure}[t]
\centering
\includegraphics[width=0.49\textwidth]{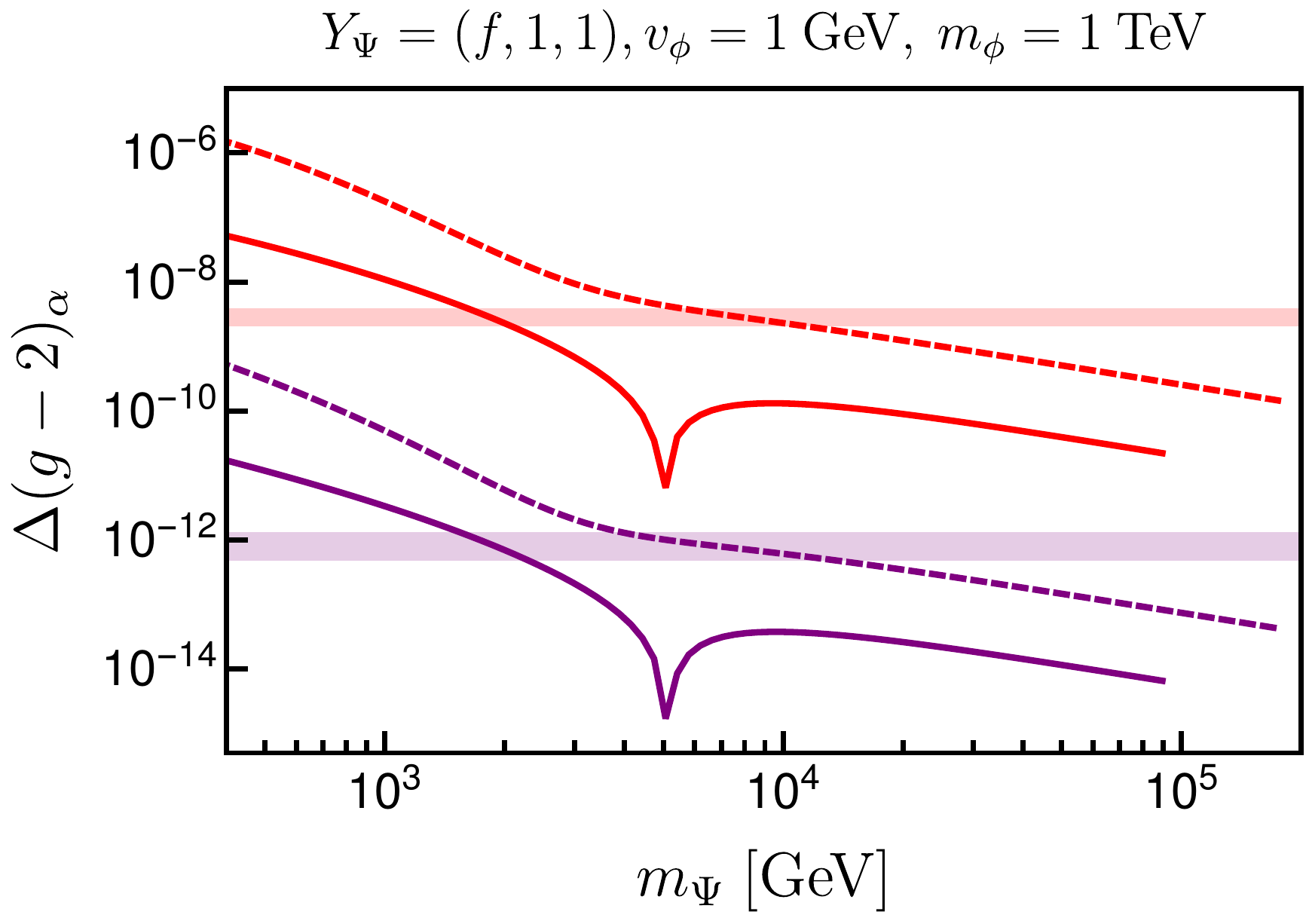}
\includegraphics[width=0.49\textwidth]{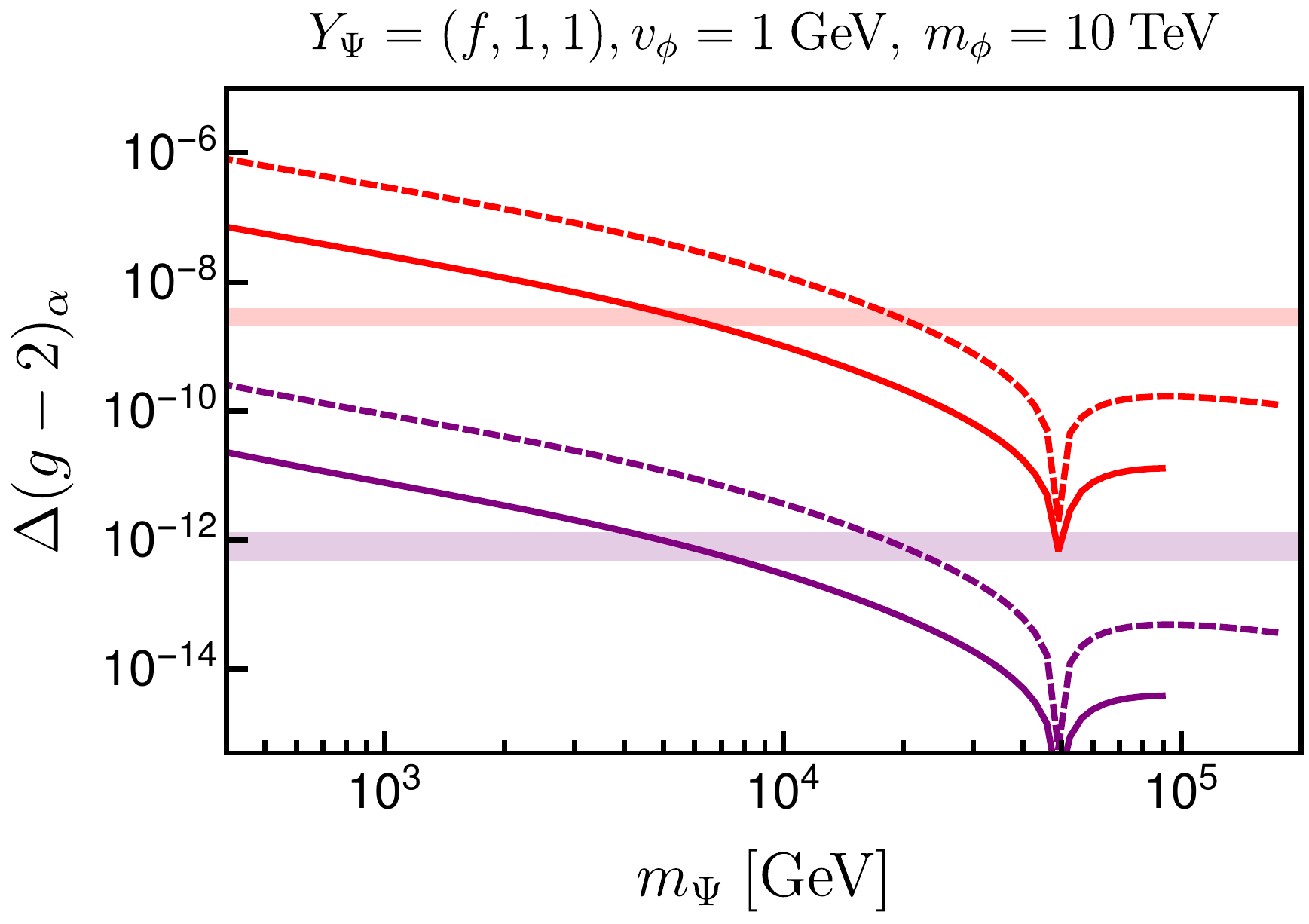}
\caption{$\Delta(g-2)_{\alpha}$, the shift with respect to the
  standard model value of $(g-2)_{\alpha}$, as a function of
  $m_{\Psi}$. The mass of the scalar $\phi$ are $m_{\phi}=1$ TeV (left) and
  $m_{\phi}=10$ TeV (right). The neutrino data is fitted with a
  diagonal $Y_{\Psi}$ and the factor $f$ has been chosen $f\simeq 0.06$
  to fit the two experimental anomalies at the same value of $m_{\Psi}$.
  This is neither necessary nor a prediction of the model and has been
  done only for demonstration. The points have been calculated assuming
  $Y_{e\phi}=Y_{e\phi^c}$ and the full dashed lines are for $(Y_{e\phi})_{ii}=1$,
  dashed lines for $(Y_{e\phi})_{ii}=4\pi$. 
      \label{BNTphigm2mPsi}
  }
\end{figure}

Since roughly $\Delta(g-2)_{\alpha} \propto 1/m_{\Psi}$,
explaining correctly the experimental anomalies would imply an upper
limit on $m_{\Psi}$. The most conservative value for this limit is
reached, if all Yukawa couplings take the maximum value allowed by
perturbativity. Taking $\forall Y \sim (4\pi)$ the result is roughly
of order $m_{\Psi} \sim {\cal O}(100)$ TeV.  This number is so
large, that it is only of academic
interest. Figure \ref{BNTphigm2mPsi} shows that for large $m_{\phi}$,
say $m_{\phi}$ larger than 1 TeV, there is practically no dependence
on the choice of $m_{\phi}$. The reason for this is that in this
limit, all the heavy scalar states decouple from the calculation and
the only contribution to the observable comes from the Goldstone
diagrams.

Figure \ref{BNTphigm2mPsi} also demonstrates, that for more reasonable
couplings again the upper limit on $m_{\Psi}$ is much lower. Allowing
$Y_{e\phi}=4 \pi$, but restricting $Y_{\Psi}$ to order ${\cal O}(1)$
the limit is roughly $(8-10)$ TeV, while for all couplings no larger
than $1$, one finds $m_{\Psi}\le (2-3)$ TeV. This last number is close
to what the LHC can probe in the high luminosity run, although the
LHC will not be able to cover the allowed range of masses completely,
see next section.

\begin{figure}[t]
\centering
\includegraphics[width=0.65\textwidth]{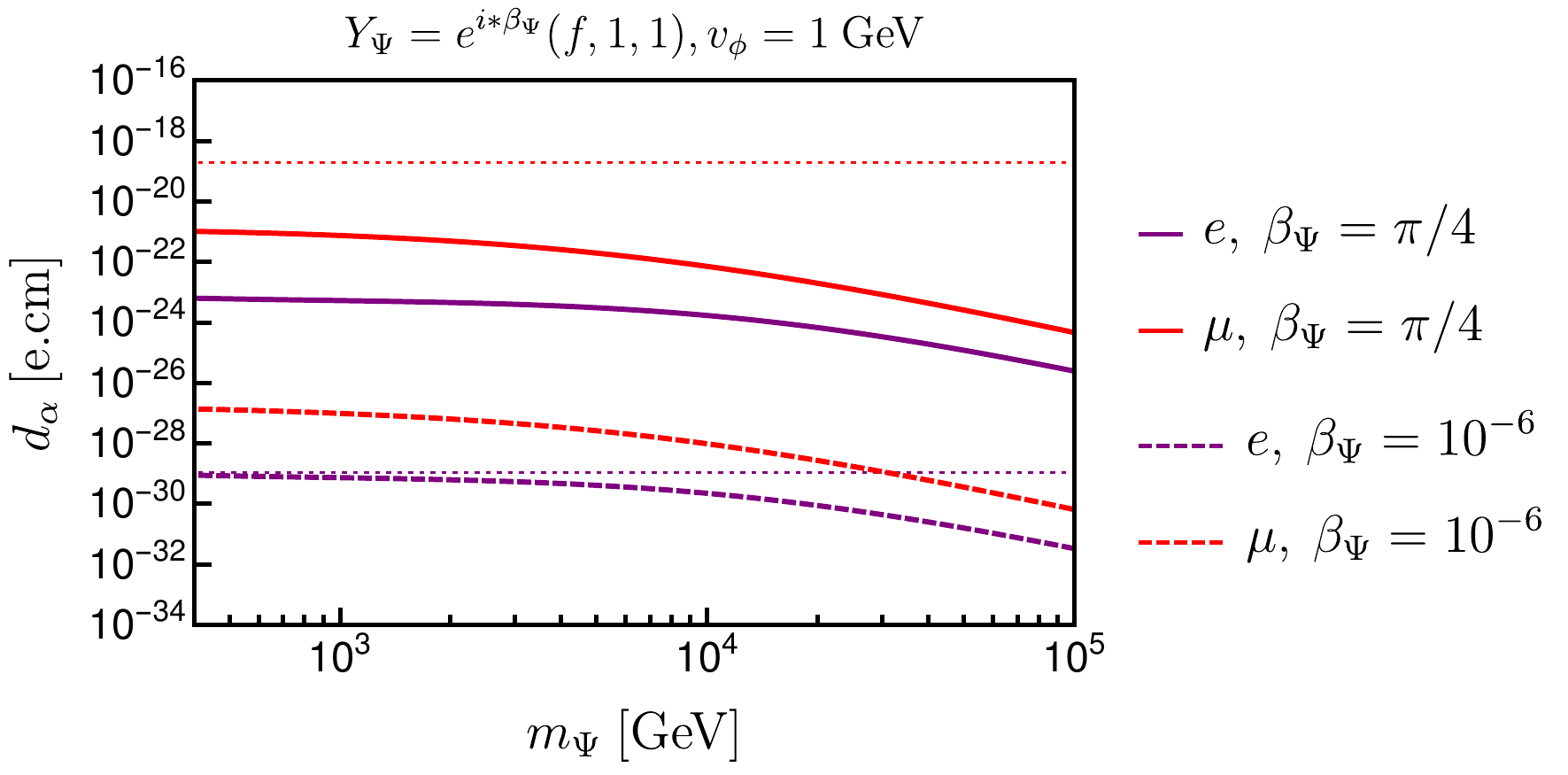}
\caption{$d_{\alpha}$, the electric dipole moments, as a function of
  $m_{\Psi}$. The large couplings, except $(Y_{\Psi})_{11}$, have all
  been taken to be equal to 1 in this plot.
      \label{BNTphiEDMmPsi}
}
\end{figure}

Let us briefly discuss the electric dipole moments, $d_{\alpha}$.
Figure \ref{BNTphiEDMmPsi} shows one example as function of
$m_{\Psi}$. The couplings $Y_{e\phi}$, $Y_{e\phi^c}$ have been taken real
and equal to one, $Y_{e\phi}=Y_{e\phi^c}=1$. The plot shows that $d_{e}$ provides a
severe constraint, while for $d_{\mu}$ there is no part of the
parameter space, where the model can saturate the experimental
bound. The plot uses the same parameters and fitting as was used in
figure \ref{BNTphigm2mPsi} for the $\Delta(g-2)_{\alpha}$. $d_{e}$
probes phases as low as $10^{-6}$. Thus, the large couplings needed to
explain $\Delta(g-2)_{e}$ essentially need to be real.

For the BNT$\phi$ model, in order to generate the CP-phase $\delta$ in
the neutrino fit, one can always put the phases into the small Yukawa
coupling. (In the case of figure \ref{BNTphiEDMmPsi} taken to be
$Y_{\overline\Psi}$). Thus, the model can survive the $d_e$ constraint
easily, but also does not make any testable predictions.

\begin{figure}[t]
\centering
\includegraphics[width=0.65\textwidth]{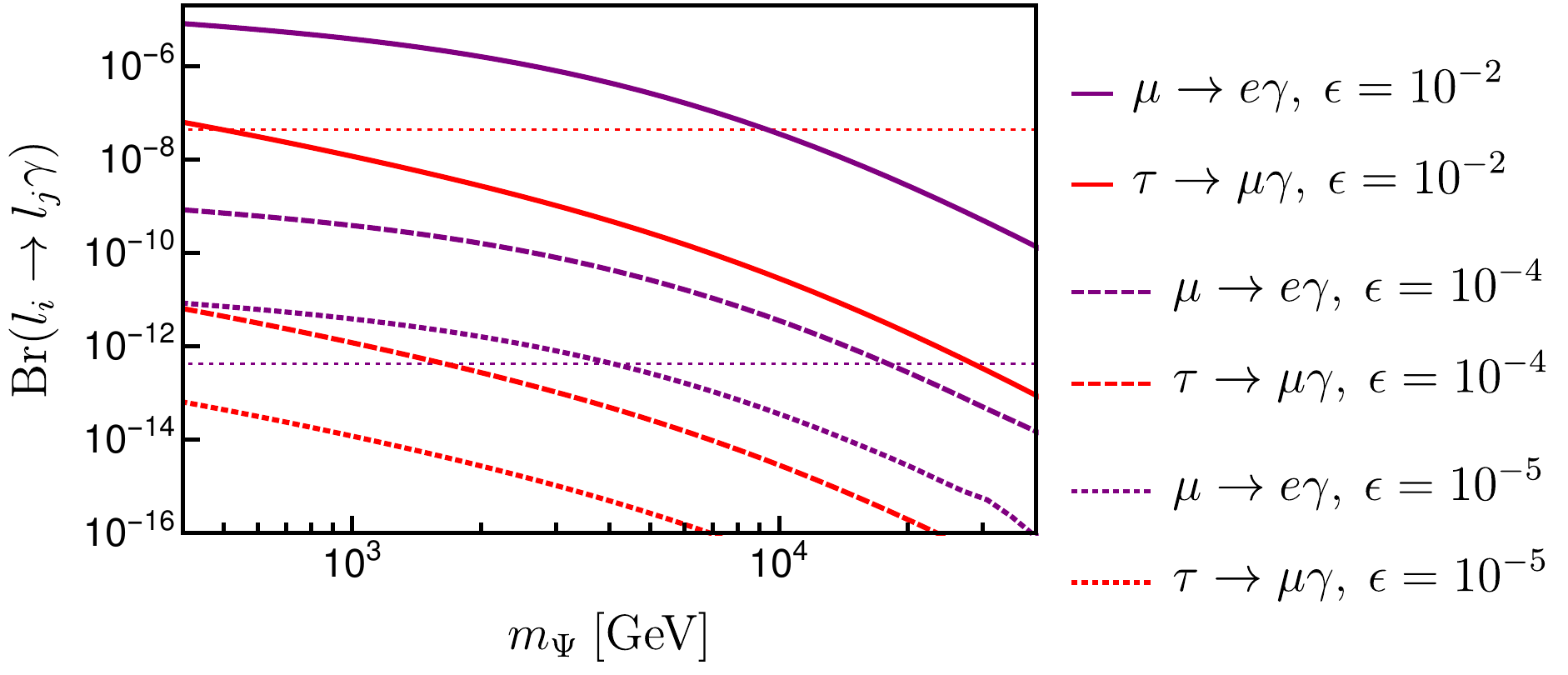}
\caption{Br$(l_i\to l_j\gamma)$ as a function of $m_{\Psi}$. The large
  couplings have all been taken to be equal to 1 in this plot.
  For the parametrization of $Y_{\Psi}$ see text. 
      \label{BNTphillgmPsi}
}
\end{figure}

The large couplings not only have to be real, they also
have to be close to diagonal, as figure \ref{BNTphillgmPsi}
demonstrates. Here, $Y_{\Psi}$ has simply be parametrized as:
\begin{equation}\label{eq:fixYPsi}
   Y_{\Psi} =
   \begin{pmatrix}
 f & \epsilon & \epsilon \\
 \epsilon & 1 & \epsilon  \\
 \epsilon & \epsilon & 1  \\
 \end{pmatrix}
\end{equation}
Here, $f\simeq 0.06$, as discussed above. $\mu\to e \gamma$ provides
a severe constraint on the off-diagonal in the $(12)$ sector:
$\epsilon \le 10^{-(5/6)}$, depending on the mass of $\Psi$.
$\tau\to\mu\gamma$ is much less stringent, but still provides
$\epsilon \le 10^{-2}$ for the lowest $m_{\Psi}$. 

In summary, the extended version of the BNT model, BNT$\phi$, can
explain the experimentally observed anomalies in $(g-2)$, while at the
same time fitting neutrino oscillation data, easily in large parts of
its parameter space. Electric dipole moments force the large Yukawa
coupling, required for $\Delta(g-2)_{\alpha}^{exp}$, to be (nearly)
real, while cLFV constrainst require them to be (nearly) diagonal.
This has some interesting consequences for the phenomenology of the
heavy fermions $\Psi$, as we are going to discuss in the next
section.

\section{Heavy fermions at colliders}
\label{sect:LHC}

\begin{figure}[t]
\centering
\includegraphics[width=0.49\textwidth]{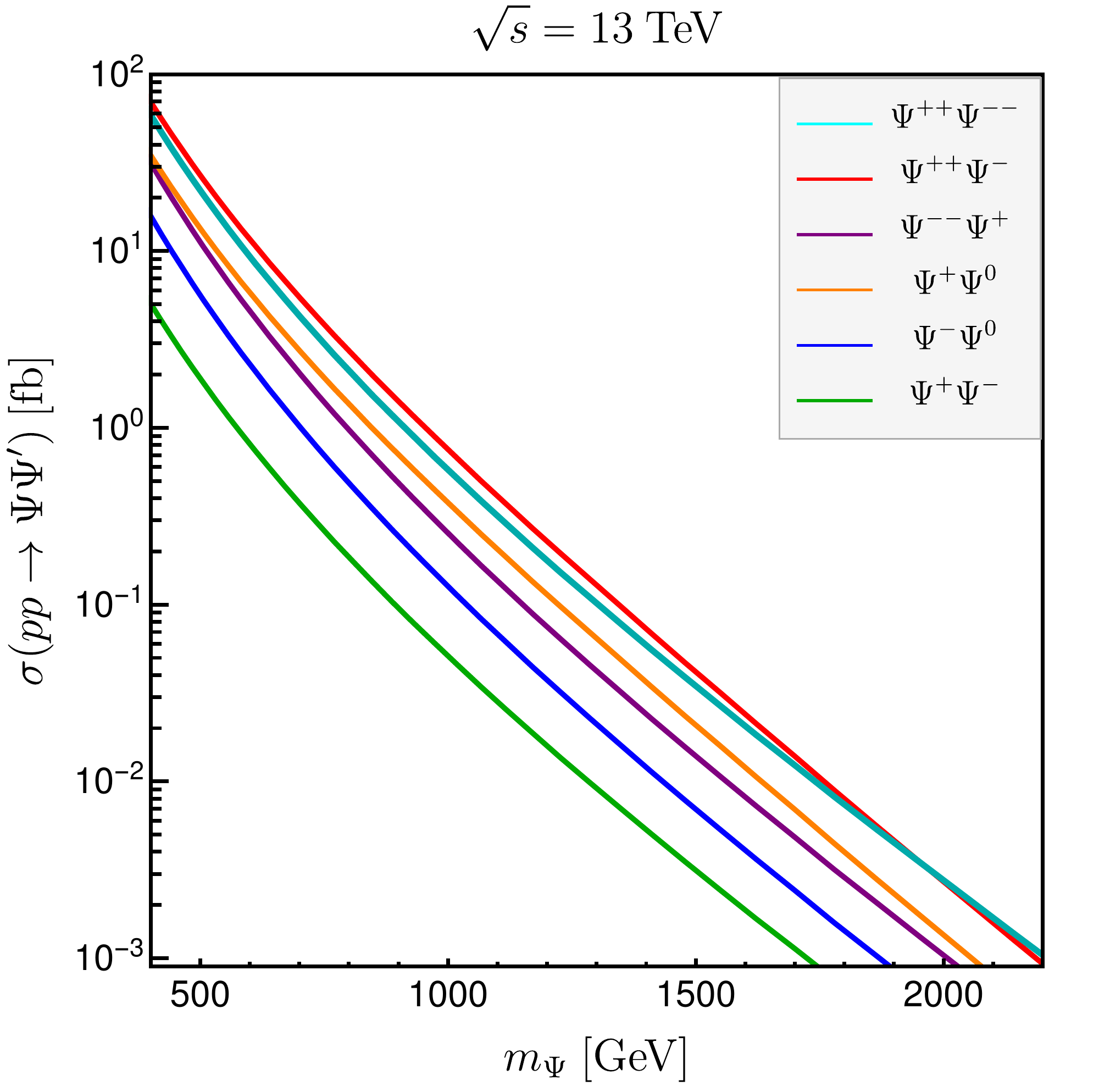}
\includegraphics[width=0.49\textwidth]{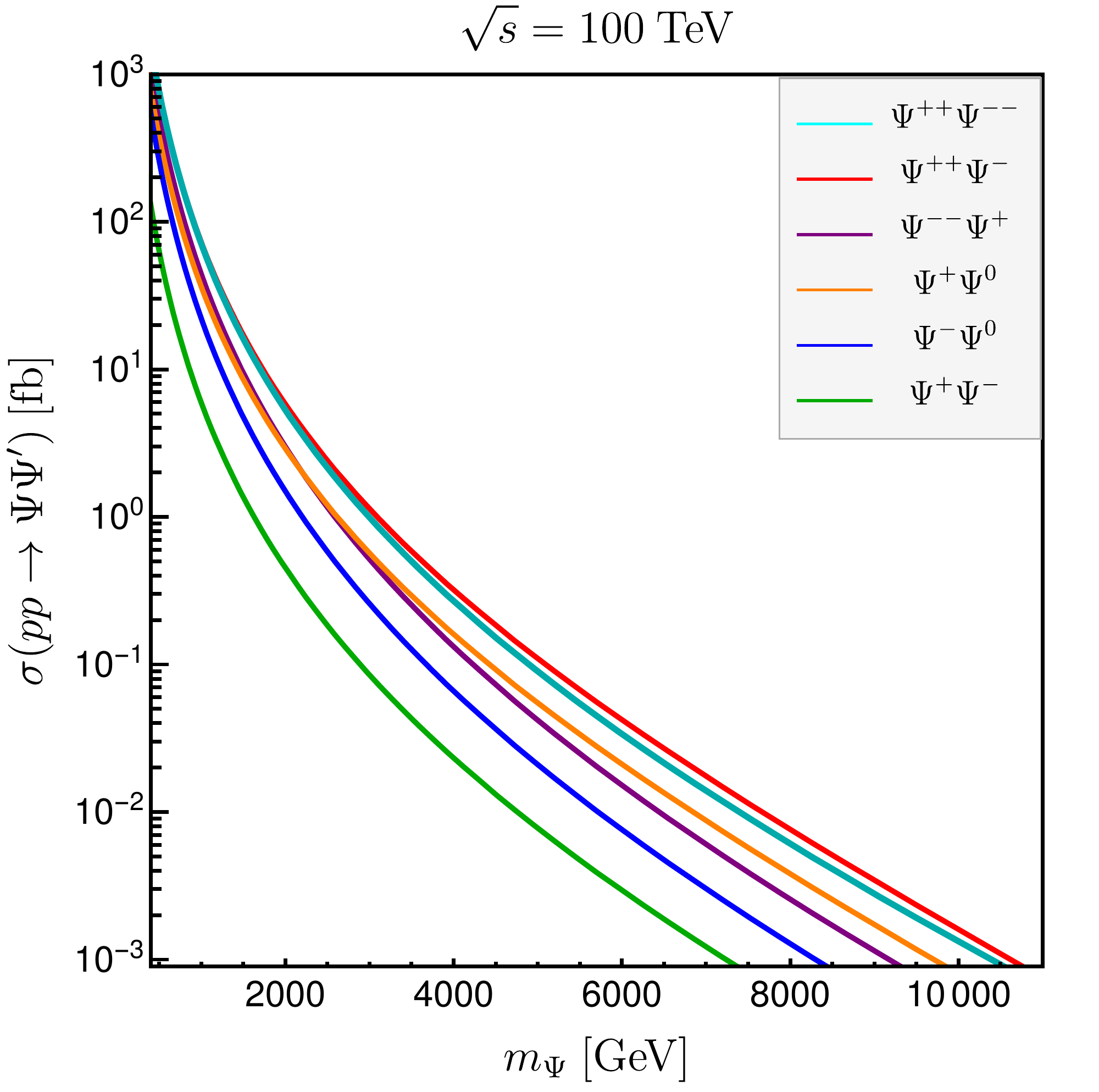}
\caption{Production cross sections for the exotic fermions in the
  BNT model for the LHC (left) and a future 100 TeV collider (right). 
  \label{XSecmPsi}
}
\end{figure}

We have calculated the production cross sections for the different
heavy fermions of the BNT model using \texttt{MadGraph}
\cite{Alwall:2007st,Alwall:2011uj,Alwall:2014hca}. For the pair
production of multiply charged particles photon-photon fusion diagrams
are especially important at large scalar masses, despite the tiny
parton density of the photon inside the proton. Manohar et
al. \cite{Manohar:2016nzj,Manohar:2017eqh} have calculated an updated
determination of the photon PDF inside the proton recently.  The
resulting \texttt{LUXqed17$\_$plus$\_$PDF4LHC15$\_$nnlo$\_$100}
combines QCD partons from \texttt{PDF4LHC15}
\cite{Butterworth:2015oua} with the LUXqed calculation of the photon
density. Results for cross sections using this set of PDFs are
shown in figure \ref{XSecmPsi} for the LHC and a hypothetical
future 100 TeV pp-collider.

The largest cross sections are pair production of the doubly charged
fermions and associated production of $\Psi^{++}\Psi^{-}$. For the
high-luminosity LHC with ${\cal L}=3/$ab each of these give more than
100 (20) events for $m_{\Psi}=1.5$ (1.8) TeV before cuts. We expect
that therefore, depending on search strategy and backgrounds, the
final reach of the LHC for discovery of the heavy fermions of the BNT
model should roughly lie in this mass range.

Possible decays of the heavy fermions depend on whether the scalars
$S$ and/or $\phi$ are lighter or heavier than $\Psi$. We will concentrate
on the case that all new scalars are heavier than the fermions for
definiteness. Final states for the different $\Psi$ are:
$\Psi^{++}_i \to l_j^+ W^+$,  $\Psi^{+}_i \to (l_j^+ Z^0,l_j^+ h^0,\nu_j+W^+)$
and $\Psi^{0}_i \to (l_j^{\pm}+W^{\mp},\nu_j +h^0,\nu_j+Z^0)$. Since
$\Psi^{0}_i$ are Majorana fermions, both lepton charges should occur
with (nearly\footnote{CP-violating phases can lead to small differences
  in the branching ratios to leptons or anti-leptons. We do not enter
  into these details, since the lepton asymmetry caused by this difference
  is a one-loop correction to the branching ratio.}) equal branching
ratios.

Before entering into a more detailed discussion of the different
branching ratios, let us comment briefly on existing limits from LHC
searches. Several ``exotic'' searches at ATLAS and CMS can provide
lower mass limits on $\Psi$. The currently most stringent one is, to
our knowledge, the multi-lepton search by CMS
\cite{Sirunyan:2019bgz}. This work uses a total of $137/$fb of
statistics to search for three charged leptons with missing energy in
the final state and no hadronic activity associated to the events.
The target process is type-III seesaw, the final state searched for
can be generated in this model via $pp\to \Sigma^+\Sigma^0 \to
(W^+\nu) + (W^+ l^-) \to l^+ l^+ l^-\nu\nu$, from the leptonic decays
of the $W$s.  For $\Sigma$ decaying ``flavour-democratically'' the
lower limit is $m_{\Sigma} =880$ GeV. (This assumes equal branching
ratios to the different lepton families. For $\Sigma$ decaying to
$\tau$'s the limit is considerably worse, see \cite{Sirunyan:2019bgz}
and for more details the earlier paper \cite{Sirunyan:2017qkz}).
While both, cross sections and branching ratios, are different in
the seesaw type-III and the BNT model, a rough estimate using
fig.(11) of \cite{Sirunyan:2019bgz} gives a lower limit on
$m_{\Psi}$ in the range of (800-900) GeV for $\Psi^{++}$ decaying
to e or $\mu$.

\begin{figure}[t]
\centering
\includegraphics[width=0.49\textwidth]{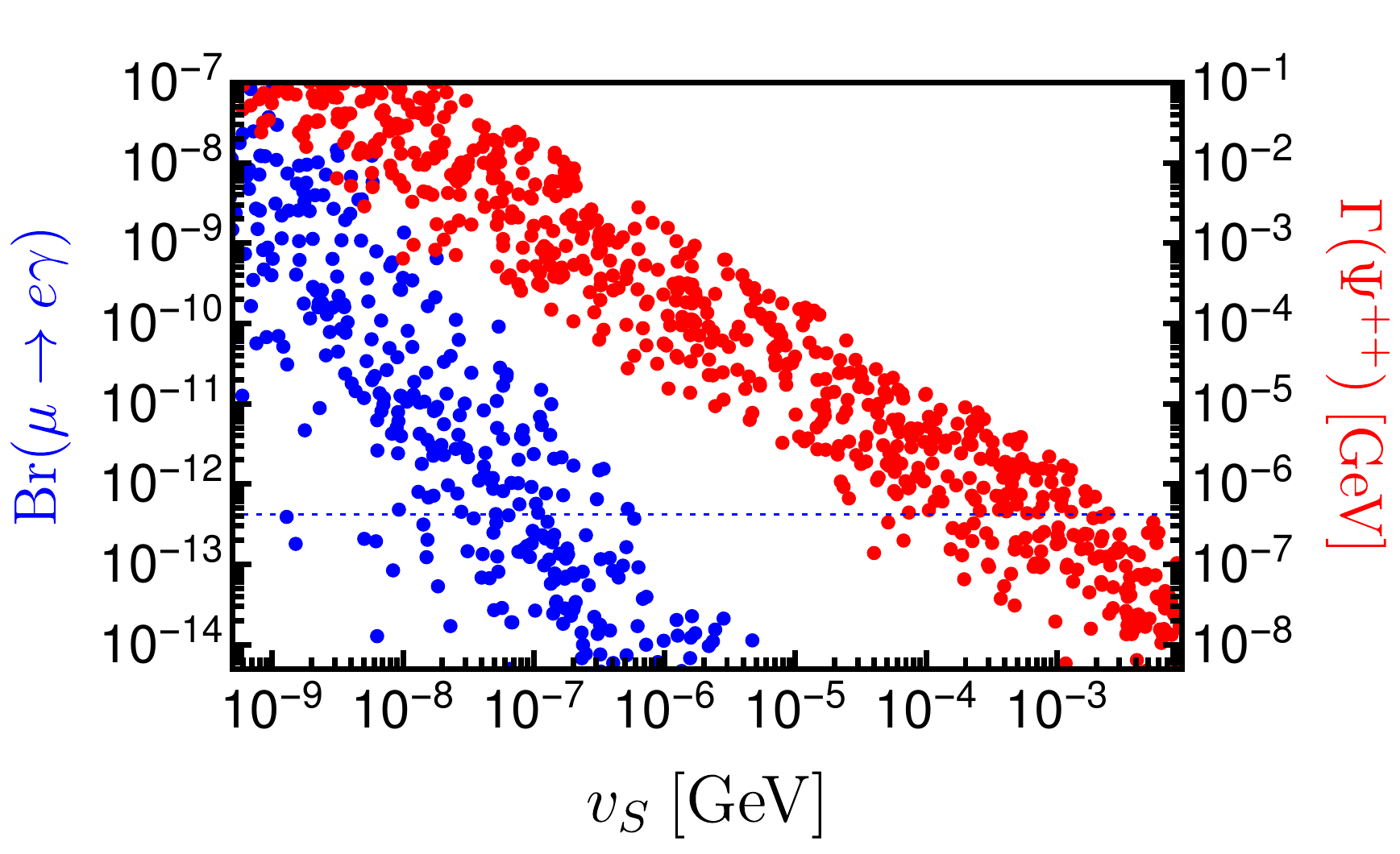}
\includegraphics[width=0.50\textwidth]{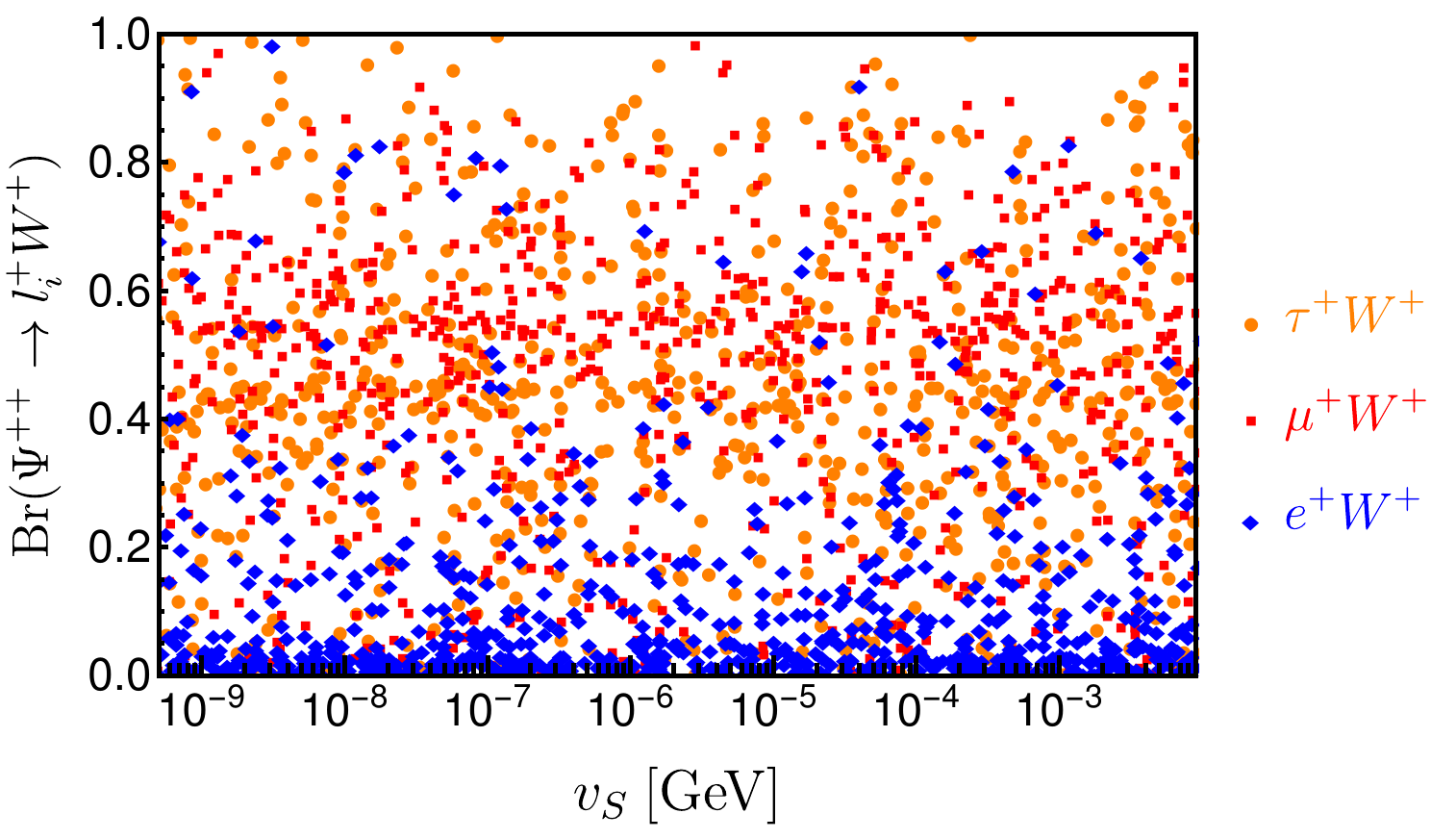}
\caption{Left: Total width of $\Psi^{++}$ and Br($\mu\to e \gamma$)
  as function of $v_S$. Right: Branching ratios of $\Psi^{++}$. The
  plot shows some random scan for a fixed mass $m_{\Psi}=1$ TeV,
  see text.
  \label{GamFpp}
}
\end{figure}

Let us turn now to a discussion of the decay branching ratios of the
heavy fermions. As discussed in the previous section, neutrino data
requires that at least one of the matrices $Y_{\Psi}$ or
$Y_{\overline \Psi}$ to be non-diagonal. Since the same couplings are
responsible for the decays of the heavy fermions, in general one
would expect that the decays of $\Psi^{++}$, $\Psi^+$ and $\Psi^0$
also violate flavour. 

An example is shown in figure \ref{GamFpp}. Here, we show the total
decay width of $\Psi^{++}$ and Br$(\mu\to e\gamma)$ (to the left) as
well as the branching ratios Br($\Psi^{++}\to l_j^+ W^+$) for
$j=e,\mu,\tau$ (right) as function of $v_S$. For this figure, we have
chosen $m_{\Psi_1}=1$ TeV and the neutrino masses were fitted with
eq. \eqref{eq:master}. In this scan we choose randomly the angles in
matrix $W$, we restricted the entries in the
matrix $T$ to be order ${\cal O}(1)$ on the diagonal and
${\cal O}(10^{-1})$ on the off-diagonal, and for simplicity $K \equiv
0$. Note that, due to the restrictions on $T$, $Y_{\Psi}$ and
$Y_{\overline \Psi}$ have a similar order of magnitude in this plot.
The plot to the left shows that both Br$(\mu\to e\gamma)$ and the
total width of $\Psi^{++}$ decrease with increasing $v_S$, since
larger $v_S$ requires smaller Yukawa couplings in the neutrino
fit. The plot to the right, however, shows no such tendency in the
branching ratios. This is easily understood: While the total width is
sensitive to the overall size of the Yukawa couplings, ratios of
branching ratios depend only on ratios of Yukawa couplings.  Thus, an
upper limit on Br$(\mu\to e\gamma)$ does not restrict the possibility
to have flavour violating $\Psi^{++}$ decays.\footnote{For flavour
  violation in heavy fermion decays one must consider the full event. For
  example, pair production of $\Psi^{++}\Psi^{--}$ leads to
  $l_i^+l_j^-+4j$ (from hadronic $W$-decays) with $i\ne j$, if
  $\Psi^{++}$ decays to more than one lepton generation.}  Results for
the decays of $\Psi^{+}$ and $\Psi^0$ show the same qualitative
behaviour, we do not repeat these plots here.

The situation is very different in those parts of parameter space,
where the model can explain $\Delta(g-2)_e$ and obey the upper bound
from cLFV decays at the same time. As discussed in the previous
section $Y_{\Psi}$ (or $Y_{\overline\Psi}$) and $Y_{e\phi}$ must be
``large'' and nearly diagonal to fit $\Delta(g-2)_e$. In this case all
heavy fermion decays are very nearly flavour diagonal.  We have
checked numerically that points with $Y_{\Psi}$ and $Y_{e\phi}$ in
the range for giving the correct  $\Delta(g-2)_e$ can not have
measurable flavour violating decays of the heavy fermions without
grossly violating existing cLFV bounds.

\begin{figure}[t]
\centering
\includegraphics[width=0.49\textwidth]{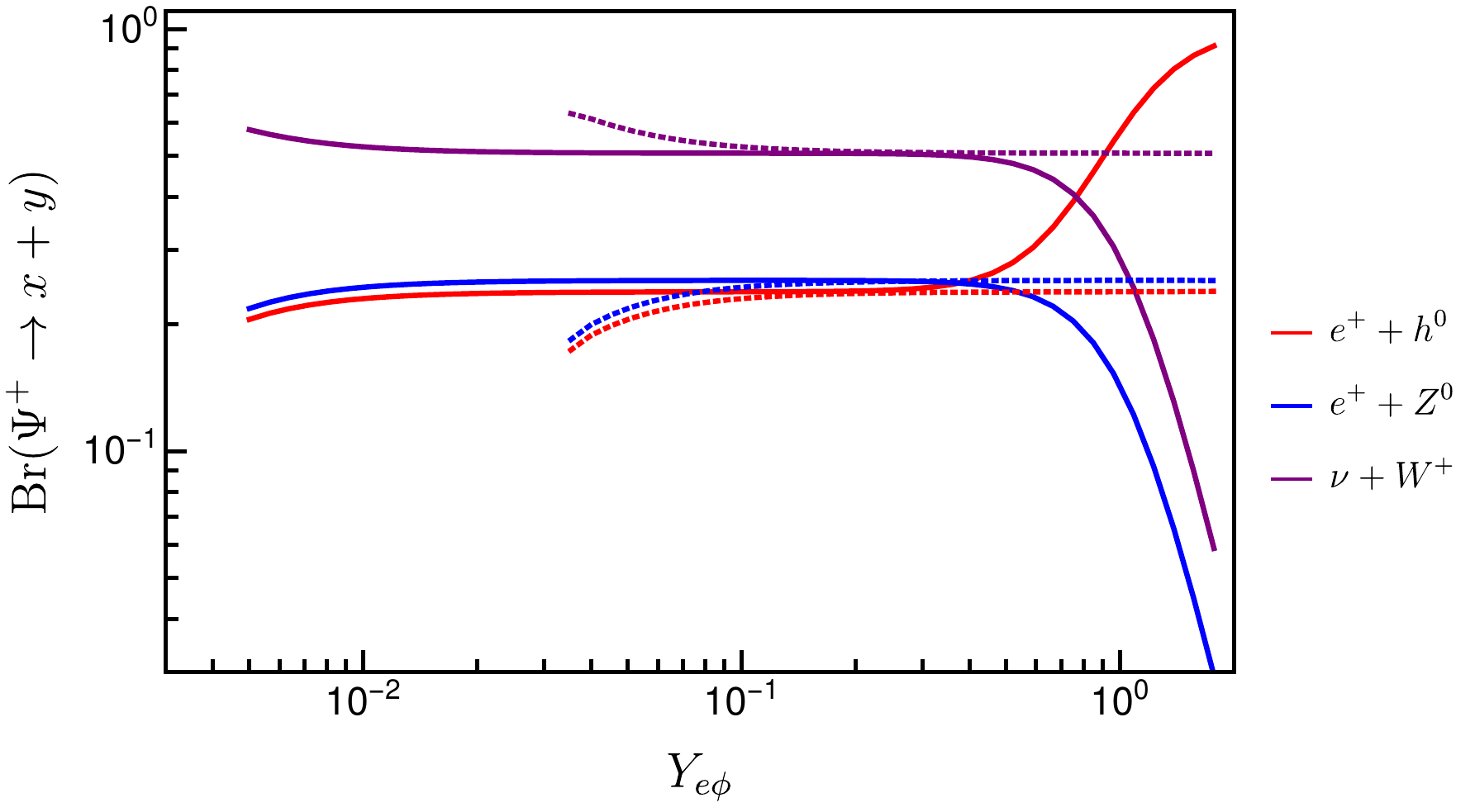}
\includegraphics[width=0.49\textwidth]{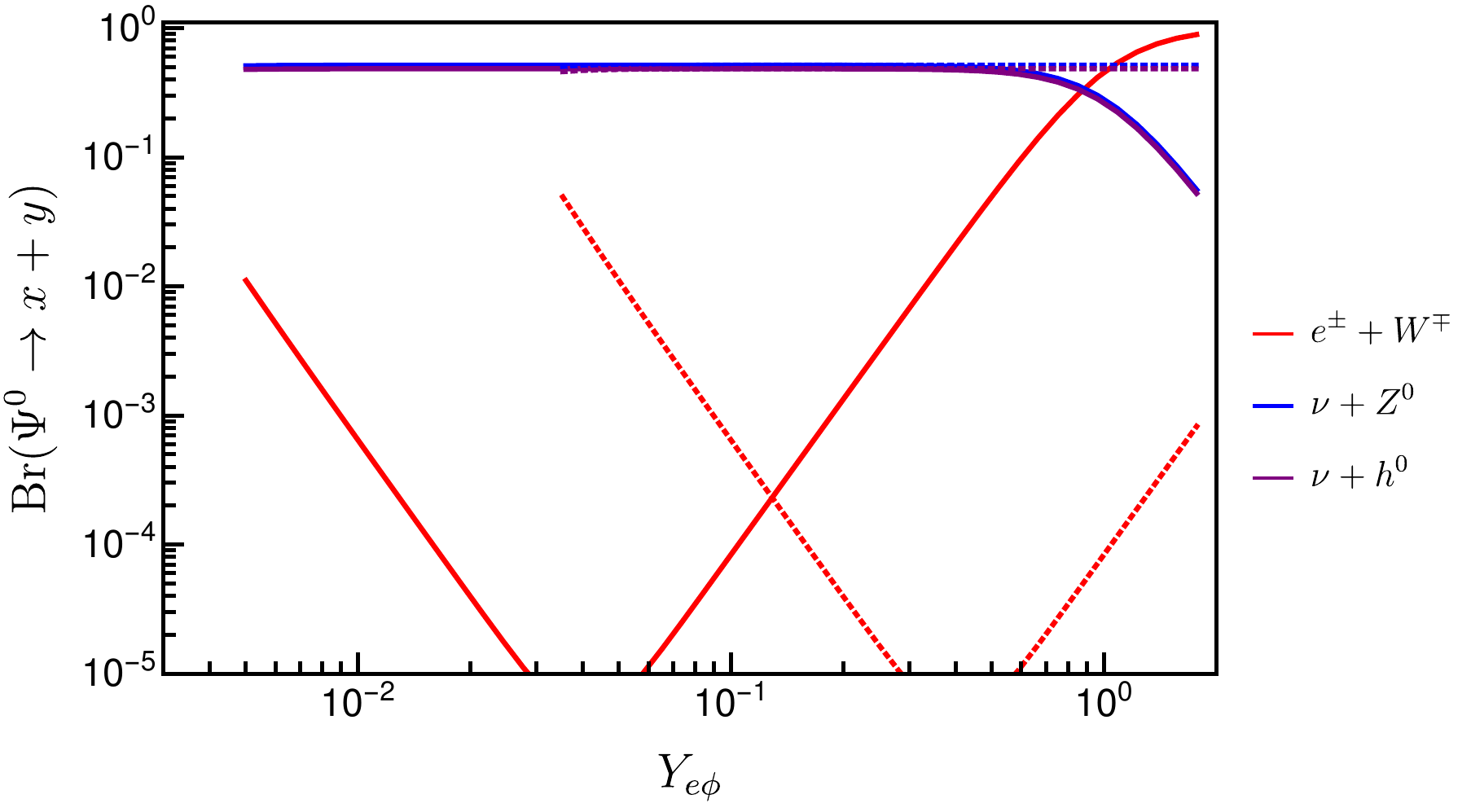}
\caption{Branching ratios for the decays of $\Psi^+_1$ and $\Psi^0_1$
  as a function of $Y_{e\phi}$, for two different values of $v_{\phi}$,
  $v_{\phi}=1$ GeV (0.1 GeV) full lines (dashed lines). All points
  in this plot are within the 1$\sigma$ c.l. range of the two
  experimental anomalies, $\Delta(g-2)_e$ and $\Delta(g-2)_{\mu}$.
  This is achieved by fitting $Y_{\Psi}$ as a function of  $Y_{e\phi}$.
  In this case, all decays are flavour-diagonal, as discussed in the text.
    \label{BrFpF0}
}
\end{figure}

The decays of $\Psi^+_1$ and $\Psi^0_1$ depend, moreover, on the value
of $v_{\phi}$ and $Y_{e\phi}$. Figure \ref{BrFpF0} shows an
example. In these plot branching ratios are shown as function of
$Y_{e\phi}$ for two values of $v_{\phi}$.  All points in this plot are
within the 1$\sigma$ c.l. range of the two experimental anomalies,
$\Delta(g-2)_e$ and $\Delta(g-2)_{\mu}$. For values of $v_{\phi} <
{\rm (few)} 10^{-3}$ GeV $\Delta(g-2)_e$ can not be fitted anymore
with perturbative Yukawa couplings. The lines stop on the left side,
when $Y_{\Psi}$ becomes non-perturbative. Note that $Y_{\Psi}$ is
fitted to the experimental data as a function of $Y_{e\phi}$ and thus,
$Y_{\Psi} >>Y_{\overline\Psi}$ in this calculation.

It is interesting to point out that for $Y_{e\phi}>>Y_{\Psi}$, which
occurs only for $v_{\phi}=1$ GeV in these figures, the decays
$\Psi^+\to e^+ h^0$ and $\Psi^0\to e^{\pm} W^{\mp}$ are enhanced.
This particular pattern appears only in the BNT$\phi$ model (and
not in the original BNT model). One can trace it back analytically
to the appearance of the coupling $Y_{e\phi}$ in the coupling of the
heavy fermions to the SM Higgs, due to the mixing of $H$ and $\phi$
(proportional to $v_{\phi}/v$). 

We note that the model does not predict the hierarchy among the
different copies of $\Psi$. Thus, the lightest of these can couple
dominantly to either $e$, $\mu$ or $\tau$.  Figure \ref{BrFpF0} shows
the case, where the lightest $\Psi$ couples to $e$. The plots for the
other cases (coupling to $\mu$ or $\tau$) are very similar and we
do not repeat them here.

In summary, the heavy fermions $\Psi$ of the BNT model can be produced
at the high-luminosity LHC. While in general one expects to have
large LFV decays of the heavy fermions, fitting $\Delta(g-2)_e$ to
the experimental anomaly, requires large and flavour-diagonal couplings.
Thus, in order for the model to explain $\Delta(g-2)_e$, the heavy
fermions must decay in a flavour conserving manner. Interestingly,
also the decays of $\Psi^+$ and $\Psi^0$ indirectly trace the
presence of $\phi$ in the model via enhanced rates for the decays 
$\Psi^+\to e^+ h^0$ and $\Psi^0\to e^{\pm} W^{\mp}$, if $Y_{e\phi}$ is
the largest Yukawa coupling. 

\section{Summary\label{sect:sum}}

We propose an extension of the original Babu-Nandi-Tavartkiladze (BNT)
neutrino model to accommodate the experimental $(g-2)_{\alpha}$
($\alpha=e, \mu$) anomalies, compatible with neutrino oscillation data
and current cLFV bounds. In this model, the presence of an extra
hypercharge zero triplet scalar gives a sizeable contribution to the
dipole operator proportional to the mass of the exotic heavy
fermions. The situation is different for the original BNT model, where
the dipole operators are suppressed by the small charged lepton
masses. Consequently, the original BNT model can not explain both
anomalies at the same time, but $\Delta(g-2)_{\mu}$ could be explained
if at least one of the exotic fermions has a mass of roughly $m_{\Psi}
\lsim (2-3)$ TeV, partially within reach of the high-luminosity LHC.

In the extended BNT model, an explanation of both observed anomalies,
$(g-2)_{e}$ and $(g-2)_{\mu}$, is compatible with neutrino
oscillation data and perturbative couplings in large part of
the available parameter space.  The smallness of the observed
$m_{\nu}$, together with the requirement to explain correctly the
$(g-2)$ anomalies, selects a specific part of the parameter space.
Specifically, $Y_{e\phi}$ and either $Y_{\Psi}$ or
$Y_{\overline{\Psi}}$ have to be order ${\cal O}(0.1-1)$. cLFV bounds
then force the large couplings to be (nearly) flavour diagonal.
Additional constraints on these couplings come from the current EDM
bounds of the electron. We showed that the experimental bound for $
\mid d_{e} \mid \sim 1.1 \times 10^{-9}e$ cm forces the large
couplings, needed to explain $\Delta(g-2)_{e}$, to be mostly real.
These results are in agreement with model-independent considerations
which can be made based on an effective field theory analysis \cite{Crivellin:2018qmi}.
For $d_{\mu}$, on the other hand, the model can not give large enough
values to saturate the current experimental bound. No observation
of $d_{\mu}$ is therefore expected in the current model.

Since cLFV current bounds require the large couplings to be nearly
diagonal in order to avoid large cLFV observables, decays of the heavy
fermions are necessarily flavour conserving. We disussed the decays of
$(\Psi^{0}, \Psi^{+}, \Psi^{++} )$ for the case that the scalars are
heavier than the fermions.  We have used \texttt{MadGraph}
\cite{Alwall:2007st,Alwall:2011uj,Alwall:2014hca} together with the
\texttt{LUXqed} PDFs \cite{Manohar:2016nzj,Manohar:2017eqh} to
calculate production cross-sections of the different heavy fermions
for the LHC and a possible future $\sqrt{s}=100$ TeV hadron collider.
For the high-luminosity LHC, future searches for the heavy fermions of
the BNT model can probe masses up to roughly $m_{\Psi} \leq (1.5-1.8)$
TeV.

We discussed branching ratios of the heavy fermions. $\Psi_{\alpha}^{++}$
decay via $\Psi_{\alpha}^{++} \rightarrow l_{\alpha}^{+}W^{+}$ with
$100\%$, whereas $\Psi_{\alpha}^{+} \rightarrow (l_{\alpha}^{+}h^{0},
l_{\alpha}^{+}Z^{0}, \nu W^{+})$. The branching ratio to
$l_{\alpha}^{+}h^{0}$ is enhanced if $Y_{e\phi}\gsim {\cal O}(1)$.
Similarly, for $\Psi^{0}$ decays the branching ratios
$\Psi_{\alpha}^{0}\rightarrow l_{\alpha}^{\pm}W^{\pm}$ can be large in
the same parts of parameter space. In summary, the extended BNT$\phi$
model can explain the observed anomalies in $(g-2)$, while making
interesting predictions for the decay patterns of the exotic
fermions.

\bigskip
\centerline{\bf Acknowledgements}

\medskip
Work supported by the Spanish grants FPA2017-85216-P
(MINECO/AEI/FEDER, UE) and PROMETEO/2018/165 grants (Generalitat
Valenciana). R.C. is also supported by FPU15/03158. R.F. acknowledges
the financial support from the Grant Agency of the Czech Republic
(GAČR) through contract number 20-17490S and from the Charles
University Research Center UNCE/SCI/013.  C.A. is supported by
FONDECYT-Chile grant No. 11180722 and ANID-Chile PIA/APOYO AFB 180002.

\bigskip

\appendix

\section{\label{sec:appendix}Appendix: Full Lagrangian expression}

The full expression of the extra interactions and mass terms of the
${\rm BNT}{\phi}$ model, including $SU(2)_{L}$ indices ($i$, $j$, ..., $p$), is as follows.

\begin{align}
\mathcal{L}_{{\rm BNT}{\phi}} & =M_{\Psi}\Psi_{ij}\overline{\Psi}_{ij}+Y_{\Psi}L_{i}\Psi_{ji}H_{j}^{*}+Y_{\overline{\Psi}}\overline{\Psi}_{lj}L_{m}S_{ijk}\epsilon_{im}\epsilon_{kl}+Y_{e\phi}e^{c}\overline{\Psi}_{ij}\phi_{ji}\nonumber \\
 & +Y_{e\phi^{c}}e^{c}\overline{\Psi}_{ij}\phi_{ij}^{*}+Y_{\Psi\phi}\Psi_{ij}\overline{\Psi}_{jk}\phi_{ki}+Y_{\Psi\phi^{c}}\Psi_{ij}\overline{\Psi}_{jk}\phi_{ik}^{*}-\mathcal{V}\,,\\
\mathcal{V} & =m_{S}^{2}S_{ijk}^{*}S_{ijk}+m_{\phi}^{2}\textrm{Tr}\left(\phi^{\dagger}\phi\right)+\left[\mu_{\phi}^{2}\textrm{Tr}\left(\phi\phi\right)+\textrm{h.c.}\right]+\left[\mu_{H\phi}H^{\dagger}\phi H\right.\nonumber \\
 & \left.\mu_{S\phi}S_{ijk}^{*}S_{ijl}\phi_{kl}+\textrm{h.c.}\right]+\frac{1}{2}\lambda_{2a}\left(S_{ijk}^{*}S_{ijk}\right)^{2}+\frac{1}{2}\lambda_{2b}S_{ijk}^{*}S_{mno}^{*}S_{ljk}S_{pno}\epsilon_{im}\epsilon_{lp}\nonumber \\
 & \frac{1}{2}\lambda_{3}\left(H^{\dagger}H\right)S_{ijk}^{*}S_{ijk}+\frac{1}{2}\lambda_{4}H_{j}H_{i}^{*}S_{klm}^{*}S_{nlm}\epsilon_{ik}\epsilon_{jn}+\lambda_{6a}\left(H^{\dagger}H\right)\textrm{Tr}\left(\phi^{\dagger}\phi\right)\nonumber \\
 & \lambda_{6b}H^{\dagger}\phi\phi^{\dagger}H+\lambda_{7a}\left(S_{ijk}^{*}S_{ijk}\right)\textrm{Tr}\left(\phi^{\dagger}\phi\right)+\lambda_{7b}S_{ijk}^{*}S_{ijn}\left(\phi\phi^{\dagger}\right)_{kn}\nonumber \\
 & \lambda_{7c}S_{ijk}^{*}S_{ilm}\left(\epsilon\phi^{*}\right)_{jk}\left(\epsilon\phi\right)_{lm}+\lambda_{8a}\left[\textrm{Tr}\left(\phi^{\dagger}\phi\right)\right]^{2}+\lambda_{8b}\textrm{Tr}\left(\phi\phi\phi^{\dagger}\phi^{\dagger}\right)\nonumber \\
 & \left\{ \lambda_{5}H_{i}H_{j}H_{k}S_{ijk}^{*}+\lambda_{9}\left(H^{\dagger}H\right)\textrm{Tr}\left(\phi\phi\right)+\lambda_{10a}S_{ijk}^{*}S_{ijk}\textrm{Tr}\left(\phi\epsilon\phi^{T}\epsilon^{T}\right)\right.\nonumber \\
 & \left.\lambda_{10b}S_{ijk}^{*}S_{ilm}\left(\epsilon\phi\epsilon^{T}\right)_{lj}\left(\epsilon\phi\epsilon^{T}\right)_{mk}+\lambda_{11}\left[\textrm{Tr}\left(\phi\phi\right)\right]^{2}+\lambda_{12}\textrm{Tr}\left(\phi\phi\right)\textrm{Tr}\left(\phi^{\dagger}\phi\right)+\textrm{h.c.}\right\} \,.
\end{align}
Whenever it was appropriate, we used a vector and matrix notation (in $SU(2)_{L}$
space) for the various fields. As usual, $\epsilon$ stands for the
2-dimensional Levi-Civita tensor.

\providecommand{\href}[2]{#2}\begingroup\raggedright\endgroup


\begin{thebibliography}{10}

\bibitem{Fukuda:1998mi}
{\bfseries Super-Kamiokande Collaboration} Collaboration, Y.~Fukuda {\em
  et~al.}, ``{Evidence for oscillation of atmospheric neutrinos},''
  \href{http://dx.doi.org/10.1103/PhysRevLett.81.1562}{{\em Phys.Rev.Lett.}
  {\bfseries 81} (1998) 1562--1567},
\href{http://arxiv.org/abs/hep-ex/9807003}{{\ttfamily arXiv:hep-ex/9807003
  [hep-ex]}}.

\bibitem{Ahmad:2002jz}
{\bfseries SNO Collaboration} Collaboration, Q.~Ahmad {\em et~al.}, ``{Direct
  evidence for neutrino flavor transformation from neutral current interactions
  in the Sudbury Neutrino Observatory},''
  \href{http://dx.doi.org/10.1103/PhysRevLett.89.011301}{{\em Phys.Rev.Lett.}
  {\bfseries 89} (2002) 011301},
\href{http://arxiv.org/abs/nucl-ex/0204008}{{\ttfamily arXiv:nucl-ex/0204008
  [nucl-ex]}}.

\bibitem{deSalas:2017kay}
P.~F. de~Salas, D.~V. Forero, C.~A. Ternes, M.~Tortola, and J.~W.~F. Valle,
  ``{Status of neutrino oscillations 2018: 3$\sigma$ hint for normal mass
  ordering and improved CP sensitivity},''
  \href{http://dx.doi.org/10.1016/j.physletb.2018.06.019}{{\em Phys. Lett.}
  {\bfseries B782} (2018) 633--640},
\href{http://arxiv.org/abs/1708.01186}{{\ttfamily arXiv:1708.01186 [hep-ph]}}.

\bibitem{deSalas:2020pgw}
P.~de~Salas, D.~Forero, S.~Gariazzo, P.~Martínez-Miravé, O.~Mena, C.~Ternes,
  M.~Tórtola, and J.~Valle, ``{2020 Global reassessment of the neutrino
  oscillation picture},'' \href{http://arxiv.org/abs/2006.11237}{{\ttfamily
  arXiv:2006.11237 [hep-ph]}}.

\bibitem{Brown:2001mga}
{\bfseries Muon g-2} Collaboration, H.~Brown {\em et~al.}, ``{Precise
  measurement of the positive muon anomalous magnetic moment},''
  \href{http://dx.doi.org/10.1103/PhysRevLett.86.2227}{{\em Phys. Rev. Lett.}
  {\bfseries 86} (2001) 2227--2231},
  \href{http://arxiv.org/abs/hep-ex/0102017}{{\ttfamily arXiv:hep-ex/0102017}}.

\bibitem{Bennett:2006fi}
{\bfseries Muon g-2} Collaboration, G.~Bennett {\em et~al.}, ``{Final Report of
  the Muon E821 Anomalous Magnetic Moment Measurement at BNL},''
  \href{http://dx.doi.org/10.1103/PhysRevD.73.072003}{{\em Phys. Rev. D}
  {\bfseries 73} (2006) 072003},
  \href{http://arxiv.org/abs/hep-ex/0602035}{{\ttfamily arXiv:hep-ex/0602035}}.

\bibitem{Jegerlehner:2009ry}
F.~Jegerlehner and A.~Nyffeler, ``{The Muon g-2},''
  \href{http://dx.doi.org/10.1016/j.physrep.2009.04.003}{{\em Phys. Rept.}
  {\bfseries 477} (2009) 1--110},
  \href{http://arxiv.org/abs/0902.3360}{{\ttfamily arXiv:0902.3360 [hep-ph]}}.

\bibitem{PDG2020}
{\bfseries Particle Data Group} Collaboration, P.~A. Zyla {\em et~al.},
  ``{Review of Particle Physics},'' {\em Prog. Theor. Exp. Phys.} {\bfseries
  2020} (2020) 083C01.

\bibitem{Keshavarzi:2018mgv}
A.~Keshavarzi, D.~Nomura, and T.~Teubner, ``{Muon $g-2$ and $\alpha(M_Z^2)$: a
  new data-based analysis},''
  \href{http://dx.doi.org/10.1103/PhysRevD.97.114025}{{\em Phys. Rev. D}
  {\bfseries 97} no.~11, (2018) 114025},
  \href{http://arxiv.org/abs/1802.02995}{{\ttfamily arXiv:1802.02995
  [hep-ph]}}.

\bibitem{Venanzoni:2014ixa}
{\bfseries Fermilab E989} Collaboration, G.~Venanzoni, ``{The New Muon g$-$2
  experiment at Fermilab},''
  \href{http://dx.doi.org/10.1016/j.nuclphysbps.2015.09.087}{{\em Nucl. Part.
  Phys. Proc.} {\bfseries 273-275} (2016) 584--588},
  \href{http://arxiv.org/abs/1411.2555}{{\ttfamily arXiv:1411.2555
  [physics.ins-det]}}.

\bibitem{Otani:2015jra}
{\bfseries E34} Collaboration, M.~Otani, ``{Status of the Muon g-2/EDM
  Experiment at J-PARC (E34)},''
  \href{http://dx.doi.org/10.7566/JPSCP.8.025008}{{\em JPS Conf. Proc.}
  {\bfseries 8} (2015) 025008}.

\bibitem{Borsanyi:2020mff}
S.~Borsanyi {\em et~al.}, ``{Leading-order hadronic vacuum polarization
  contribution to the muon magnetic momentfrom lattice QCD},''
  \href{http://arxiv.org/abs/2002.12347}{{\ttfamily arXiv:2002.12347
  [hep-lat]}}.

\bibitem{Crivellin:2020zul}
A.~Crivellin, M.~Hoferichter, C.~A. Manzari, and M.~Montull, ``{Hadronic vacuum
  polarization: $(g-2)_\mu$ versus global electroweak fits},''
  \href{http://arxiv.org/abs/2003.04886}{{\ttfamily arXiv:2003.04886
  [hep-ph]}}.

\bibitem{Parker:2018vye}
R.~H. Parker, C.~Yu, W.~Zhong, B.~Estey, and H.~Müller, ``{Measurement of the
  fine-structure constant as a test of the Standard Model},''
  \href{http://dx.doi.org/10.1126/science.aap7706}{{\em Science} {\bfseries
  360} (2018) 191}, \href{http://arxiv.org/abs/1812.04130}{{\ttfamily
  arXiv:1812.04130 [physics.atom-ph]}}.

\bibitem{Aoyama:2017uqe}
T.~Aoyama, T.~Kinoshita, and M.~Nio, ``{Revised and Improved Value of the QED
  Tenth-Order Electron Anomalous Magnetic Moment},''
  \href{http://dx.doi.org/10.1103/PhysRevD.97.036001}{{\em Phys. Rev. D}
  {\bfseries 97} no.~3, (2018) 036001},
  \href{http://arxiv.org/abs/1712.06060}{{\ttfamily arXiv:1712.06060
  [hep-ph]}}.

\bibitem{CarcamoHernandez:2019ydc}
A.~Cárcamo~Hernández, S.~King, H.~Lee, and S.~Rowley, ``{Is it possible to
  explain the muon and electron $g-2$ in a $Z'$ model?},''
  \href{http://dx.doi.org/10.1103/PhysRevD.101.115016}{{\em Phys. Rev. D}
  {\bfseries 101} no.~11, (2020) 11},
  \href{http://arxiv.org/abs/1910.10734}{{\ttfamily arXiv:1910.10734
  [hep-ph]}}.

\bibitem{Crivellin:2018qmi}
A.~Crivellin, M.~Hoferichter, and P.~Schmidt-Wellenburg, ``{Combined
  explanations of $(g-2)_{\mu,e}$ and implications for a large muon EDM},''
  \href{http://dx.doi.org/10.1103/PhysRevD.98.113002}{{\em Phys. Rev.}
  {\bfseries D98} no.~11, (2018) 113002},
\href{http://arxiv.org/abs/1807.11484}{{\ttfamily arXiv:1807.11484 [hep-ph]}}.

\bibitem{Dorsner:2020aaz}
I.~Do\v{r}sner, S.~Fajfer, and S.~Saad, ``{$\mu \to e \gamma$ selecting scalar
  leptoquark solutions for the $(g-2)_{e,\mu}$ puzzles},''
  \href{http://arxiv.org/abs/2006.11624}{{\ttfamily arXiv:2006.11624
  [hep-ph]}}.

\bibitem{Liu:2018xkx}
J.~Liu, C.~E. Wagner, and X.-P. Wang, ``{A light complex scalar for the
  electron and muon anomalous magnetic moments},''
  \href{http://dx.doi.org/10.1007/JHEP03(2019)008}{{\em JHEP} {\bfseries 03}
  (2019) 008}, \href{http://arxiv.org/abs/1810.11028}{{\ttfamily
  arXiv:1810.11028 [hep-ph]}}.

\bibitem{Dutta:2018fge}
B.~Dutta and Y.~Mimura, ``{Electron $g-2$ with flavor violation in MSSM},''
  \href{http://dx.doi.org/10.1016/j.physletb.2018.12.070}{{\em Phys. Lett. B}
  {\bfseries 790} (2019) 563--567},
  \href{http://arxiv.org/abs/1811.10209}{{\ttfamily arXiv:1811.10209
  [hep-ph]}}.

\bibitem{Bauer:2019gfk}
M.~Bauer, M.~Neubert, S.~Renner, M.~Schnubel, and A.~Thamm, ``{Axion-like
  particles, lepton-flavor violation and a new explanation of $a_\mu$ and
  $a_e$},'' \href{http://dx.doi.org/10.1103/PhysRevLett.124.211803}{{\em Phys.
  Rev. Lett.} {\bfseries 124} no.~21, (2020) 211803},
  \href{http://arxiv.org/abs/1908.00008}{{\ttfamily arXiv:1908.00008
  [hep-ph]}}.

\bibitem{Hiller:2019mou}
G.~Hiller, C.~Hormigos-Feliu, D.~F. Litim, and T.~Steudtner, ``{Anomalous
  magnetic moments from asymptotic safety},''
  \href{http://arxiv.org/abs/1910.14062}{{\ttfamily arXiv:1910.14062
  [hep-ph]}}.

\bibitem{CarcamoHernandez:2020pxw}
A.~Cárcamo~Hernández, Y.~Hidalgo~Velásquez, S.~Kovalenko, H.~Long, N.~A.
  Pérez-Julve, and V.~Vien, ``{Fermion spectrum and $g-2$ anomalies in a low
  scale 3-3-1 model},'' \href{http://arxiv.org/abs/2002.07347}{{\ttfamily
  arXiv:2002.07347 [hep-ph]}}.

\bibitem{Hati:2020fzp}
C.~Hati, J.~Kriewald, J.~Orloff, and A.~Teixeira, ``{Anomalies in $^8$Be
  nuclear transitions and $(g-2)_{e,\mu}$: towards a minimal combined
  explanation},'' \href{http://arxiv.org/abs/2005.00028}{{\ttfamily
  arXiv:2005.00028 [hep-ph]}}.

\bibitem{Chen:2020jvl}
C.-H. Chen and T.~Nomura, ``{Electron and muon $g-2$, radiative neutrino mass,
  and $\ell' \to \ell \gamma$ in a $U(1)_{e-\mu}$ model},''
  \href{http://arxiv.org/abs/2003.07638}{{\ttfamily arXiv:2003.07638
  [hep-ph]}}.

\bibitem{Calibbi:2020emz}
L.~Calibbi, M.~López-Ibáñez, A.~Melis, and O.~Vives, ``{Muon and electron
  $g-2$ and lepton masses in flavor models},''
  \href{http://dx.doi.org/10.1007/JHEP06(2020)087}{{\em JHEP} {\bfseries 06}
  (2020) 087}, \href{http://arxiv.org/abs/2003.06633}{{\ttfamily
  arXiv:2003.06633 [hep-ph]}}.

\bibitem{Botella:2020xzf}
F.~J. Botella, F.~Cornet-Gomez, and M.~Nebot, ``{Electron and muon $g-2$
  anomalies in general flavour conserving two Higgs doublets models},''
  \href{http://arxiv.org/abs/2006.01934}{{\ttfamily arXiv:2006.01934
  [hep-ph]}}.

\bibitem{Chen:2020tfr}
K.-F. Chen, C.-W. Chiang, and K.~Yagyu, ``{An explanation for the muon and
  electron $g-2$ anomalies and dark matter},''
  \href{http://arxiv.org/abs/2006.07929}{{\ttfamily arXiv:2006.07929
  [hep-ph]}}.

\bibitem{Dutta:2020scq}
B.~Dutta, S.~Ghosh, and T.~Li, ``{Explaining $(g-2)_{\mu,e}$, KOTO anomaly and
  MiniBooNE excess in an extended Higgs model with sterile neutrinos},''
  \href{http://arxiv.org/abs/2006.01319}{{\ttfamily arXiv:2006.01319
  [hep-ph]}}.

\bibitem{Jana:2020pxx}
S.~Jana, V.~P. K., and S.~Saad, ``{Resolving electron and muon $g-2$ within the
  2HDM},'' \href{http://dx.doi.org/10.1103/PhysRevD.101.115037}{{\em Phys. Rev.
  D} {\bfseries 101} no.~11, (2020) 115037},
  \href{http://arxiv.org/abs/2003.03386}{{\ttfamily arXiv:2003.03386
  [hep-ph]}}.

\bibitem{Babu:2009aq}
K.~S. Babu, S.~Nandi, and Z.~Tavartkiladze, ``{New Mechanism for Neutrino Mass
  Generation and Triply Charged Higgs Bosons at the LHC},''
  \href{http://dx.doi.org/10.1103/PhysRevD.80.071702}{{\em Phys. Rev.}
  {\bfseries D80} (2009) 071702},
\href{http://arxiv.org/abs/0905.2710}{{\ttfamily arXiv:0905.2710 [hep-ph]}}.

\bibitem{Staub:2012pb}
F.~Staub, ``{SARAH 3.2: Dirac Gauginos, UFO output, and more},''
  \href{http://dx.doi.org/10.1016/j.cpc.2013.02.019}{{\em Comput. Phys.
  Commun.} {\bfseries 184} (2013) 1792--1809},
\href{http://arxiv.org/abs/1207.0906}{{\ttfamily arXiv:1207.0906 [hep-ph]}}.

\bibitem{Staub:2013tta}
F.~Staub, ``{SARAH 4 : A tool for (not only SUSY) model builders},''
  \href{http://dx.doi.org/10.1016/j.cpc.2014.02.018}{{\em Comput. Phys.
  Commun.} {\bfseries 185} (2014) 1773--1790},
\href{http://arxiv.org/abs/1309.7223}{{\ttfamily arXiv:1309.7223 [hep-ph]}}.

\bibitem{Porod:2003um}
W.~Porod, ``{SPheno, a program for calculating supersymmetric spectra, SUSY
  particle decays and SUSY particle production at e+ e- colliders},''
  \href{http://dx.doi.org/10.1016/S0010-4655(03)00222-4}{{\em Comput. Phys.
  Commun.} {\bfseries 153} (2003) 275--315},
\href{http://arxiv.org/abs/hep-ph/0301101}{{\ttfamily arXiv:hep-ph/0301101
  [hep-ph]}}.

\bibitem{Porod:2011nf}
W.~Porod and F.~Staub, ``{SPheno 3.1: Extensions including flavour, CP-phases
  and models beyond the MSSM},''
  \href{http://dx.doi.org/10.1016/j.cpc.2012.05.021}{{\em Comput. Phys.
  Commun.} {\bfseries 183} (2012) 2458--2469},
\href{http://arxiv.org/abs/1104.1573}{{\ttfamily arXiv:1104.1573 [hep-ph]}}.

\bibitem{Cordero-Carrion:2018xre}
I.~Cordero-Carrión, M.~Hirsch, and A.~Vicente, ``{Master Majorana neutrino
  mass parametrization},''
  \href{http://dx.doi.org/10.1103/PhysRevD.99.075019}{{\em Phys. Rev. D}
  {\bfseries 99} no.~7, (2019) 075019},
  \href{http://arxiv.org/abs/1812.03896}{{\ttfamily arXiv:1812.03896
  [hep-ph]}}.

\bibitem{Cordero-Carrion:2019qtu}
I.~Cordero-Carrión, M.~Hirsch, and A.~Vicente, ``{General parametrization of
  Majorana neutrino mass models},''
  \href{http://dx.doi.org/10.1103/PhysRevD.101.075032}{{\em Phys. Rev. D}
  {\bfseries 101} no.~7, (2020) 075032},
  \href{http://arxiv.org/abs/1912.08858}{{\ttfamily arXiv:1912.08858
  [hep-ph]}}.

\bibitem{Alwall:2007st}
J.~Alwall, P.~Demin, S.~de~Visscher, R.~Frederix, M.~Herquet, F.~Maltoni,
  T.~Plehn, D.~L. Rainwater, and T.~Stelzer, ``{MadGraph/MadEvent v4: The New
  Web Generation},''
  \href{http://dx.doi.org/10.1088/1126-6708/2007/09/028}{{\em JHEP} {\bfseries
  09} (2007) 028},
\href{http://arxiv.org/abs/0706.2334}{{\ttfamily arXiv:0706.2334 [hep-ph]}}.

\bibitem{Alwall:2011uj}
J.~Alwall, M.~Herquet, F.~Maltoni, O.~Mattelaer, and T.~Stelzer, ``{MadGraph 5
  : Going Beyond},'' \href{http://dx.doi.org/10.1007/JHEP06(2011)128}{{\em
  JHEP} {\bfseries 1106} (2011) 128},
\href{http://arxiv.org/abs/1106.0522}{{\ttfamily arXiv:1106.0522 [hep-ph]}}.

\bibitem{Alwall:2014hca}
J.~Alwall, R.~Frederix, S.~Frixione, V.~Hirschi, F.~Maltoni, O.~Mattelaer,
  H.~S. Shao, T.~Stelzer, P.~Torrielli, and M.~Zaro, ``{The automated
  computation of tree-level and next-to-leading order differential cross
  sections, and their matching to parton shower simulations},''
  \href{http://dx.doi.org/10.1007/JHEP07(2014)079}{{\em JHEP} {\bfseries 07}
  (2014) 079},
\href{http://arxiv.org/abs/1405.0301}{{\ttfamily arXiv:1405.0301 [hep-ph]}}.

\bibitem{Manohar:2016nzj}
A.~Manohar, P.~Nason, G.~P. Salam, and G.~Zanderighi, ``{How bright is the
  proton? A precise determination of the photon parton distribution
  function},'' \href{http://dx.doi.org/10.1103/PhysRevLett.117.242002}{{\em
  Phys. Rev. Lett.} {\bfseries 117} no.~24, (2016) 242002},
\href{http://arxiv.org/abs/1607.04266}{{\ttfamily arXiv:1607.04266 [hep-ph]}}.

\bibitem{Manohar:2017eqh}
A.~V. Manohar, P.~Nason, G.~P. Salam, and G.~Zanderighi, ``{The Photon Content
  of the Proton},'' \href{http://dx.doi.org/10.1007/JHEP12(2017)046}{{\em JHEP}
  {\bfseries 12} (2017) 046},
\href{http://arxiv.org/abs/1708.01256}{{\ttfamily arXiv:1708.01256 [hep-ph]}}.

\bibitem{Butterworth:2015oua}
J.~Butterworth {\em et~al.}, ``{PDF4LHC recommendations for LHC Run II},''
  \href{http://dx.doi.org/10.1088/0954-3899/43/2/023001}{{\em J. Phys.}
  {\bfseries G43} (2016) 023001},
\href{http://arxiv.org/abs/1510.03865}{{\ttfamily arXiv:1510.03865 [hep-ph]}}.

\bibitem{Sirunyan:2019bgz}
{\bfseries CMS} Collaboration, A.~M. Sirunyan {\em et~al.}, ``{Search for
  physics beyond the standard model in multilepton final states in
  proton-proton collisions at $\sqrt{s} =$ 13 TeV},''
  \href{http://dx.doi.org/10.1007/JHEP03(2020)051}{{\em JHEP} {\bfseries 03}
  (2020) 051}, \href{http://arxiv.org/abs/1911.04968}{{\ttfamily
  arXiv:1911.04968 [hep-ex]}}.

\bibitem{Sirunyan:2017qkz}
{\bfseries CMS} Collaboration, A.~M. Sirunyan {\em et~al.}, ``{Search for
  Evidence of the Type-III Seesaw Mechanism in Multilepton Final States in
  Proton-Proton Collisions at $\sqrt{s}=13\text{ }\text{ }\mathrm{TeV}$},''
  \href{http://dx.doi.org/10.1103/PhysRevLett.119.221802}{{\em Phys. Rev.
  Lett.} {\bfseries 119} no.~22, (2017) 221802},
  \href{http://arxiv.org/abs/1708.07962}{{\ttfamily arXiv:1708.07962
  [hep-ex]}}.

\end{thebibliography}
\end{document}